\documentclass[a4paper,11pt]{article}
\usepackage[utf8]{inputenc}
\usepackage{jheppub}
\usepackage{amsmath,amsfonts,amssymb}
\usepackage{textcomp}
\usepackage{graphicx}
\usepackage{bm}
\usepackage{color}
\usepackage{verbatim}
\usepackage{bbold}
\usepackage{slashed}
\usepackage{subcaption}
\usepackage{epsfig}
\usepackage{hyperref}

\newcommand{\p}{\partial}

\newcommand{\rar}{\rightarrow}

\newcommand{\ut}{\dot{t}}
\newcommand{\ux}{\dot{x}}
\newcommand{\uy}{\dot{y}}
\newcommand{\uz}{\dot{z}}
\newcommand{\ur}{\dot{r}}
\newcommand{\M}{\mathcal{M}}
\newcommand{\N}{\mathcal{N}}

\title{Holographic fermions at strong translational symmetry breaking: a Bianchi-VII case study}

\author[b]{A. Bagrov,}
\emailAdd{abagrov@science.ru.nl}
\author[a,1]{N. Kaplis,}
\note{https://orcid.org/0000-0002-4425-3240}
\emailAdd{kaplis@lorentz.leidenuniv.nl}
\author[a,c,2]{A. Krikun,}
\note{https://orcid.org/0000-0001-8789-8703}
\emailAdd{krikun@lorentz.leidenuniv.nl}
\author[a]{K. Schalm,}
\emailAdd{kschalm@lorentz.leidenuniv.nl}
\author[a]{and J. Zaanen}
\emailAdd{zaanen@lorentz.leidenuniv.nl}

\affiliation[a]{Institute Lorentz $\Delta$ITP, Leiden University, \\ PO Box 9506, Leiden 2300 RA, The Netherlands}
\affiliation[b]{Institute for Molecules and Materials, Radboud University, \\ Heyendaalseweg 135, Nijmegen 6525 AJ, The Netherlands}
\affiliation[c]{Institute for Theoretical and Experimental Physics (ITEP)\footnote{On leave from}, \\ B. Cheryomushkinskaya 25, 117218 Moscow, Russia }

\abstract{
It is presently unknown how strong lattice potentials influence the fermion spectral function of the holographic strange metals predicted by the AdS/CFT correspondence. This embodies a crucial test for the application of holography to condensed matter experiments. We show that for one particular momentum direction this spectrum can be computed for arbitrary strength of the effective translational symmetry breaking potential of the so-called Bianchi-VII geometry employing ordinary differential equations. Deep in the strange metal regime we find rather small changes to the single-fermion response computed by the emergent quantum critical IR, even when the potential becomes relevant in the infra-red. However, in the regime where holographic quasi-particles occur, defining a Fermi surface in the continuum, they acquire a finite lifetime at any finite potential strength. At the transition from irrelevancy to relevancy of the Bianchi potential in the deep infra-red the quasi-particle remnants disappear completely and the fermion spectrum exhibits a purely relaxational behaviour.
}

\preprint{}

\begin{document}
\maketitle

\section{Introduction}
\label{sec:intro}

The holographic description of strongly interacting quantum systems at finite density \cite{Zaanen:2015oix} has proven very helpful in offering new explanations for the mysterious nature of the electron systems formed in high-$T_c$ superconductors and other strongly correlated fermion systems \cite{Keimer2015}. However, a systematic theoretical understanding of the equilibrium physics is only available in the spatial continuum. In the gravitational bulk one can use the homogeneity of the (stationary) space-time in order to write the equations of motions simply as ordinary differential equations (ODE's) involving only the radial coordinate of the emergent extra holographic dimension. The relevant laboratory systems are characterized by very strong lattice potentials breaking translational symmetry \cite{Keimer2015}. These lattices alter the nature of the IR physics, and although it is well understood how to incorporate translational symmetry breaking (TSB) in holography \cite{Horowitz:2012ky, Horowitz:2012gs, Blake:2013owa}, they greatly complicate the bulk computations. In such geometries one resorts to numerical solutions of systems of non-linear partial differential equations (PDE's). Of course this can be and has been done, but presently only a limited number of results are available, dealing with simple bulk systems (typically Einstein-Maxwell) and describing mainly macroscopic transport properties \cite{Horowitz:2012ky, Rangamani:2015hka}. 

More specifically, most computations focus on the (optical) electrical and thermal conductivities at zero momentum. This is partly because of the relative ease with which these can be calculated -- the equations simplify greatly by considering only zero momentum, and even more so by considering the zero frequency (DC) limit. The price to be paid for this convenience is that these macroscopic conductivities contain relatively limited information regarding the system. In this limit hydrodynamics becomes dominant, i.e. any finite density system will turn into a perfect metal and the conductivities are to first order governed by momentum relaxation due to the breaking of translational symmetry; this is rather insensitive to the detailed nature of the system and/or the actual pattern of TSB. As a consequence one can get quite far using simplified holographic models like massive gravity \cite{Vegh:2013sk} or Q-lattices/axions \cite{Andrade:2013gsa} that introduce momentum non-conservation without an explicit lattice. Since these simplified models do not capture the physics of Umklapp they have no bearing on the behaviour of physical quantities at microscopic momenta. 

Transport, i.e. conductivity at zero momentum, plays actually a relatively minor role in the large range of properties that are measured in the condensed matter laboratories, where its constraints have been realized all along. There is much to be learned by looking at finite momenta. This is already informative for linear response, either probed through electromagnetic means \cite{PhysRevB.90.035143}, through inelastic neutron scattering, which gives access to the full momentum and energy dependence of the dynamical magnetic susceptibility \cite{Keimer2015}, through the novel electron energy-loss instruments \cite{2015arXiv150904230V} that promise to give access to the full dynamical charge susceptibility in the near future, and especially through the mature photo-emission and scanning tunnelling spectroscopy techniques that deliver direct information regarding the single electron propagators \cite{Zaanen:2015oix}.

Although holographic photo-emission \cite{Cubrovic:2009ye,Faulkner:2010zz} has played a key role in the early history of AdS/CMT, there have been surprisingly few studies on holographic spectral functions in the presence of lattices \cite{Liu:2012tr, Ling:2013aya,Ling:2014bda}, and these only focus on the cases where the lattice deformation is irrelevant in the infra-red. Here we go beyond this, to study fermion spectral function at strong lattice potentials. This difficult question is precisely one where the virtues of the holographic approach come into play. The holographic set-up is the so-called \hbox{Bianchi-VII} (helical) background \cite{Donos:2012js}. This is a member of the ``homogeneous lattice'' family of geometries that have been used in holography in order to simulate  effects of translational symmetry breaking \cite{Donos:2011ff,Donos:2012wi,Donos:2013eha}. Compared to an actual lattice these models have enhanced symmetries allowing one to use the geometry's homogeneity in order to easily compute its properties using just ODE's even in the regimes where the TSB deformation is strong. In the context of the Bianchi-VII background this applies when the probe momentum lies on the direction of the helix. For generic momenta, which are more natural since they give rise to Umklapp effects, the usual TSB intuition applies and one has to deal with PDE's.

\begin{figure}[th]
\centering
\begin{subfigure}{0.45\linewidth}
\center
\includegraphics[width=1.\linewidth]{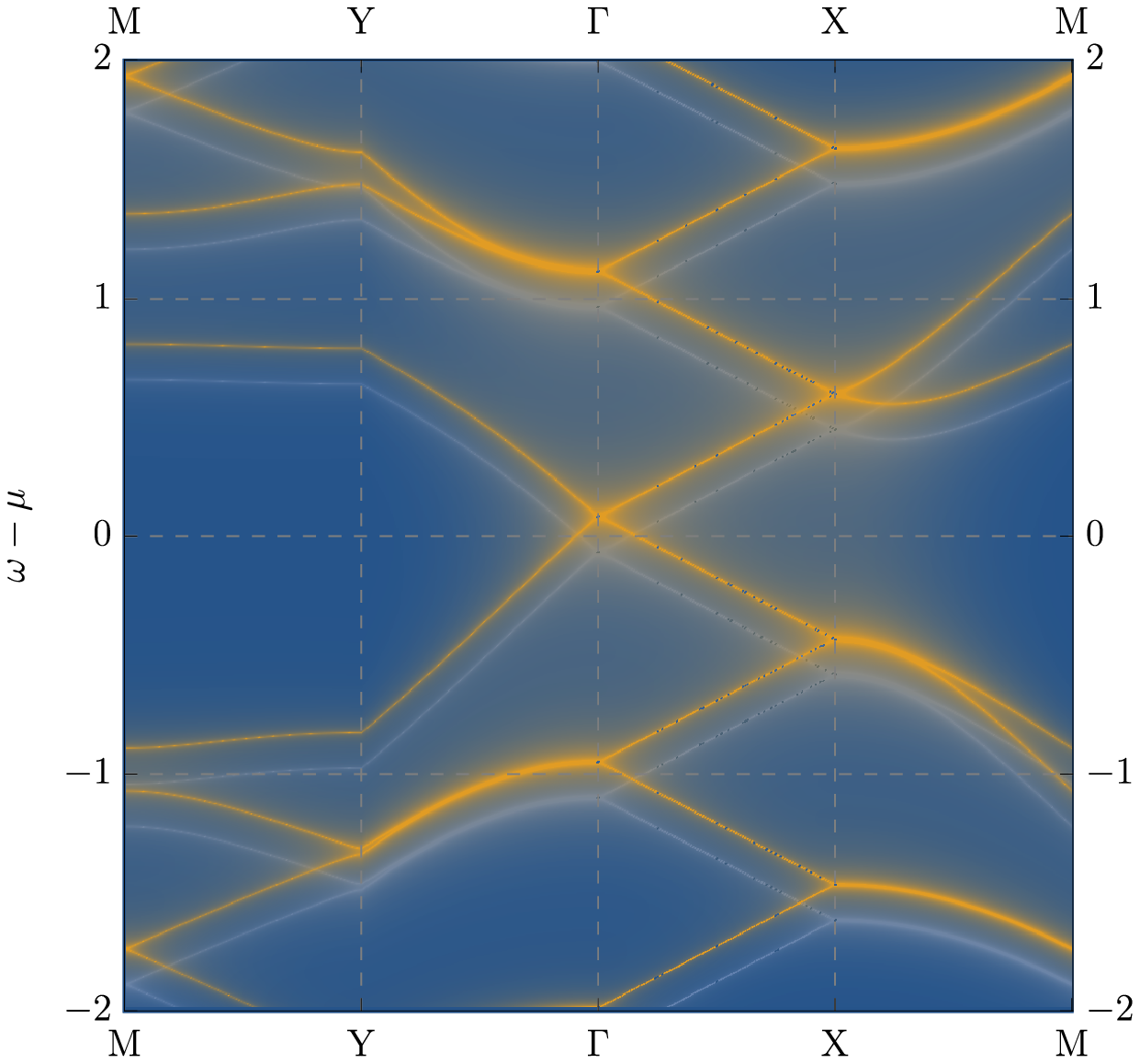}
\caption{Cartoon of the band structure of boundary fermion. The $k_\perp$-dependent Umklapp gaps are apparent in the $\mathrm{X}-\mathrm{M}$ and $\mathrm{Y}-\mathrm{\Gamma}$ cuts. Different colours refer to opposite chiralities.}
\label{fig:bands}
\end{subfigure}
~
\begin{subfigure}{0.45\linewidth}
\center
\includegraphics[width=0.7\linewidth]{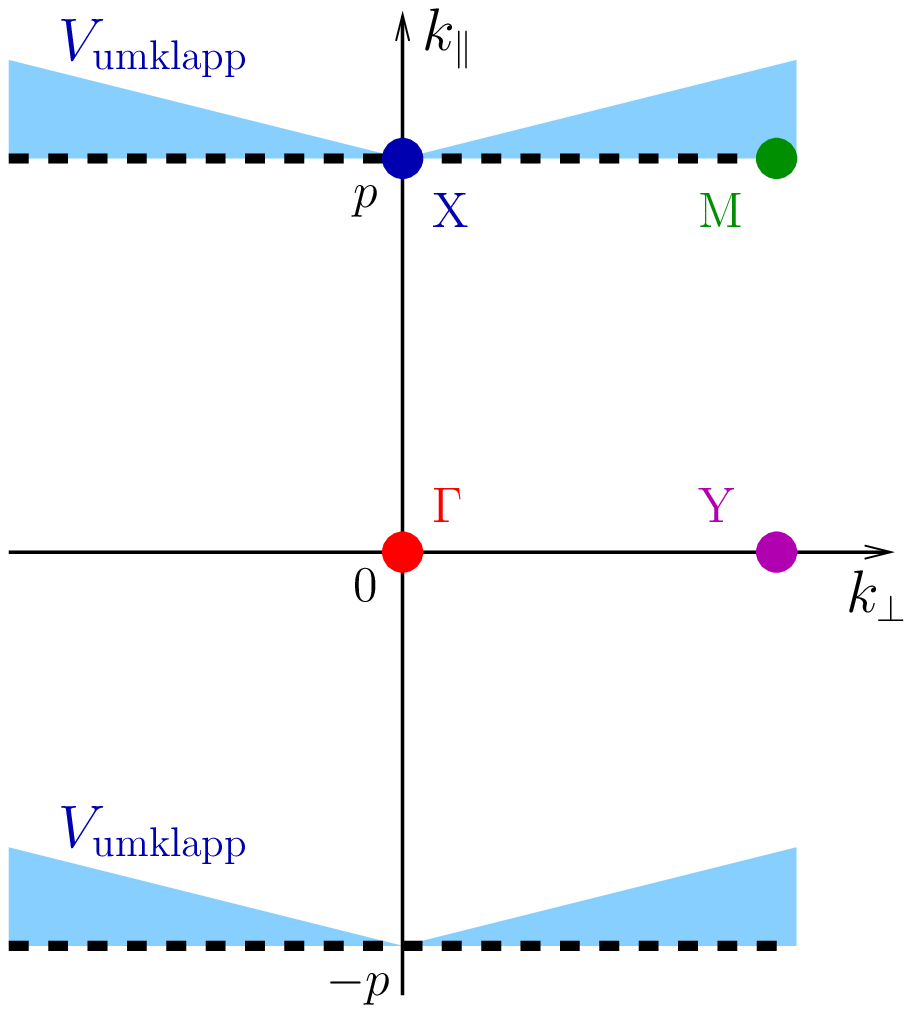}
\vspace{0.4 cm}
\caption{2D Brillouin zone, Umklapp surfaces are shown as thick dashed lines. The profile of the Umklapp coupling is shown in blue.}
\label{fig:BrilZone}
\end{subfigure}
\caption{Phenomenology of the effective boundary theory.}
\end{figure}

We will consider single-fermion two-point functions of the dual boundary QFT in this holographic background, but it is quite useful to first put our work into context by considering the features of these fermions in the effective potential corresponding to the Bianchi-VII geometry. Even though superficially this helix-like potential resembles that of helical magnetic order (of the kind encountered in MnSi \cite{Muhlbauer915}) it is actually quite different, due to the peculiar behaviour of the Umklapp scattering. As we will show in detail in Sec.~\ref{sec:Dirac_equation}, the fermionic ``boundary'' degrees of freedom we study holographically are $3+1$-dim Weyl fermions $\theta$. From the holographic theory and its effective action \eqref{eq:boundary_action} we can obtain the equations of motion of the free Weyl fermions in the presence of the potential induced by a helical source with amplitude $\lambda$, pitch $p$ and direction along the $O_x$-axis. It reads
\begin{gather}
\label{eq:boundary_equation}
 \Big[ \p_\mu \gamma^\mu + \left(\mu  + \mu^5 \gamma^5 \right) \gamma^0 + V_{\mathrm{Umklapp}} \Big] \theta = 0, \\
 \notag
V_{\mathrm{Umklapp}} \sim   \lambda^2 \Big[ \cos(2 p x) (k_y \gamma^y - k_z \gamma^z) - \sin(2 p x) ( k_z  \gamma^y - k_y \gamma^z ) \Big], \\ \notag
\mu^5 \sim \lambda^2.
\end{gather}
The helical deformation breaks chiral symmetry and this results in a constant shift of the effective chemical potential of left and right Weyl fermions at all momenta, similar to the Stoner (spin) splitting found in simple ferromagnetic metals. Moreover the spatial dependence of the helical source leads to a uni-directional breaking of translations with a periodicity set by its pitch. This implies a Brillouin zone indicated in Fig.\,\ref{fig:BrilZone} in terms of the parallel $k_{\parallel} \sim k_x$ and perpendicular $k_{\perp} \sim k_y, k_z$ momenta. The Umklapp scattering has the effect of mixing the states of opposite spin of the chiral Weyl fermions and because the potential is harmonic this involves a momentum transfer $k_x \rightarrow k_x + 2p$. The interesting part is that the specific Bianchi-VII construction dictates that the Umklapp coupling is proportional to $k_\perp$, i.e. it vanishes as long as $k_y = k_z = 0$ \eqref{eq:boundary_equation}: this property is the reason why for $k_{\perp} = 0$ the problem reduces to a simpler ODE system. Based on standard band structure wisdom this is already counter-intuitive since in the holographic gravitational bulk standard lattice potentials would have the effect of maximizing Umklapp in this direction. In our case it is by moving away from this direction that the Umklapp switches on, causing band gaps at high symmetry points as one can see on the cartoon Fig.\,\ref{fig:bands}: one can clearly see the absence of Umklapp in the $\Gamma-X$ direction ($k_\perp=0$) and the gaps appearing in the $X-M$ and $\Gamma-Y$ cuts.\footnote{At the $Y$-point there is no splitting in the first band, which has $k_{\parallel} = 0$, because it corresponds to the flat wave in the periodic direction, hence the spatial modulation is averaged out. Nevertheless the Umklapp at the $Y$-point is observed in the second band, which originates from the hybridization of the nested fermionic dispersions with relative momentum $k_\perp + 4 p$.}

What this simplified band structure misses is the life-time of the boundary fermions. In the effective action \eqref{eq:boundary_equation} this is simply ignored and these quasi-particles are infinitely long-lived. The true holographic model is a fully consistent and interacting theory, however. Indeed, these theories famously include a non-trivial IR critical sector possibly interacting with a more conventional protected mode, such as a Goldstone boson or a Fermi-surface excitation. The Umklapp described above is an additional universal feature on top of these interactions. To illustrate the effects of this most clearly, we restrict ourselves to $k_\perp=0$ and by making use of the simplifications discussed above we will study in detail the effects of translational symmetry breaking on the finite momentum fermionic response, in the presence of interactions. 

A particular aspect we focus on is that the holographic model under consideration undergoes a quantum phase transition where the Bianchi-VII TSB potential turns from irrelevant to relevant in the IR, as a function of the pitch and strength of the helical potential (Fig.\,\ref{fig:phase_diagram})\cite{Donos:2012js}. Accordingly, the optical conductivity in the parallel direction turns from relevant to irrelevant upon crossing the quantum phase transition. In both cases $\sigma_\parallel (\omega) \sim \omega^{\alpha}$ and at the transition the exponent turns from negative to positive, respectively. This was called a ``metal-insulator'' transition, but this is a confusing terminology. In condensed matter physics an insulator invariably refers to an incompressible state of matter, characterized by an energy scale below which conduction is suppressed exponentially. The relevant TSB regime of the Bianchi-VII model describes a state that is still compressible and subject to continuous renormalisation while the current operator just becomes irrelevant. This is alien to finite-density, free-fermion physics of conventional condensed matter, but it is a distinct possibility in an interacting Quantum Field Theory. For instance, consider the (engineering) scaling of the conductivity in a $d+1$ dimensional CFT $\sigma (\omega) \sim \omega^{d-2}$ which for $d \ge 3$ turns irrelevant. As we will see soon, probing the system with fermions (instead of currents) reveals that the physics has no relation to a conventional metal-insulator distinction. In the remainder we will therefore call the two phases of the Bianchi-VII system the irrelevant and relevant translation symmetry breaking (TSB) phases.

Let us briefly summarize our findings. As we discuss in Sec.\,\ref{sec:analytic}, the dynamics and the spectrum of the fermionic excitations can be neatly understood in terms of the semi-analytic framework developed for the Reissner-Nordstr"om (RN) metal in \cite{Faulkner:2009wj, Faulkner:2011tm}. The Dirac equation in the holographic gravitational bulk can be recast as a Schr\"odinger equation and the corresponding potential (Fig.\,\ref{fig:potentials_finite_omega}) reveals the qualitative features of the fermionic response. It can either have well defined quasi-particles, or be dominated by diffusive dynamics, or be completely governed by the strange metallic ``un-particle'' Green function. When we consider the effect of the Bianchi-VII TSB deformation we can distinguish two scenarios depending on the scaling dimension of the fermionic operator under consideration. 

For fermions with low close-to-free field scaling dimension, for which sharp quasi-particle peaks are observed at finite chemical potential (dual to a pristine RN black hole), an effect is seen even in the irrelevant TSB phase for any finite helical amplitude $\lambda$. The barrier of the Schr\"odinger potential that encodes the IR physics becomes finite (Fig.\,\ref{fig:potentials_finite_omega}) and the fermion has a finite tunnelling probability to reach the horizon, leading to the quasi-particles acquiring a finite life-time at zero energy. The Fermi-surface becomes smeared accordingly. Notice that this is in striking contrast with the principles governing real Fermi-liquids in a periodic potential.

Computing the actual spectral function of the dual CFT this effect is also clearly seen as we show in Sec.\,\ref{sec:spectral_function}. Upon increasing the potential strength the broadening continues to increase (Figs.~\ref{fig:DensityCuts}, \ref{fig:FermiPoleVsLambda}). At the same time a qualitative change appears to take place at the transition to the relevant TSB phase. Instead of the damped but still recognizable quasi-particle spectra (left panel on Fig.\,\ref{fig:spectral_met-ins}) of the irrelevant case, a completely relaxational response is found (right panel on Fig.\,\ref{fig:spectral_met-ins}). To better understand what is going on, it is informative to track the spectrum of the AdS bulk quasi-normal modes in the complex plane, which we do in Sec.\,\ref{sec:quasinormal_modes} (Figs.~\ref{fig:DensityQNMs}, \ref{fig:QNMs_analisys}). These are in one-to-one correspondence with the poles of the fermionic Green's function in the dual field theory. As a function of increasing $\lambda$ one finds that the quasi-particle poles not only reveal the increasing damping but also a decrease of the Fermi-velocity indicating that the mass of the quasi-particles is increasing. At the transition from irrelevant to relevant TSB we find that the poles corresponding to the quasi-particles disappear, ``dissolving'' in the string of thermal poles on the imaginary axis that describe the ``un-particle'' $\mathcal{G}$ branch cuts.

The other scenario where the scaling dimension of the fermionic operator is large is arguably most revealing since the boundary UV fermions directly probe the deep infra-red, being unimpeded by the quasi-particle resonance. The effects of the Bianchi deformation are less spectacular in this regime -- in stark contrast with the quasi-particle regime there is still a hard potential wall in IR (Fig.\,\ref{fig:potentials_finite_omega}) resulting in the spectral weight vanishing identically at precisely zero energy and temperature. Only quantitative changes occur as the strength of the TSB is increased including the transition from irrelevancy to relevancy (Fig.\,\ref{fig:density_cuts}). 

We will conclude by analysing these results in the context of the particular way translational symmetry is broken in this model (Sec.\,\ref{sec:conclusion}). The main text is followed by three Appendixes which are devoted to some details about the numerics (App.\,\ref{sec:AppendixQNMs}), the small-$\omega$ IR Green's function (App.\,\ref{app:IR_green}) and the special case of zero frequency Schr\"odinger potential (App.\,\ref{app:zero_frequency}).
\section{The Bianchi-VII background as a model of TSB in holography}
\label{sec:helical_background}
In \cite{Donos:2012js} the Bianchi-VII homogeneous space-time with a helical symmetry was suggested as a toy holographic model for studying effects of translational symmetry breaking and physics of metal-insulator transition. The following discussion will be based on this set-up, so here we recall its structure.

In order to explicitly break the translational symmetry one introduces an additional (massive) vector field on top of the usual Einstein-Maxwell action in 5D AdS bulk:
\begin{gather}
\label{eq:action_helix}
\mathcal{S} = \int dx^5 \, \sqrt{-g} \left( R + 12 -\frac{1}{4}  F_{\M\N}F^{\M\N} - \frac{1}{4}W_{\M\N}W^{\M\N} - \frac{m^2_B}{2}B_{\M} B^{\M} \right), 
\end{gather}
where $F=dA$ is the strength of the Maxwell field $A_\M$, $m_B$ is a mass of the vector field $B_\M$, $W=dB$ its strength, and we set the curvature radius to unity and $m_B$ to zero\footnote{In \cite{Donos:2012js} the additional Chern-Simons term $\mathcal{L}_{CS} \sim B \wedge F \wedge W$ was present but we discard it here because it is not strictly necessary to generate the helical background.}. The 5D spatial indices are $\M, \N \in \{t,x,y,z, r\}$, the boundary 4D theory is spanned by the usual Minkowski coordinates $\mu,\nu \in \{t,x,y,z\}$ and is located at radial infinity $r \rar \infty$ in our notation. While the asymptotic value of $A$ at $r\rar \infty$ defines the boundary chemical potential $\mu$, the boundary value of $B$ is set by the spatially dependent helical ``source'' $\varLambda(x)$, which breaks translation symmetry.
\begin{equation}
\label{eq:source_B}
 A(r)\Big|_{r\rar \infty} = \mu dt, \qquad B(r,x)\Big|_{r\rar \infty} = \varLambda(x) \equiv \lambda \omega_2,
\end{equation}
where the helical 1-forms $\omega_i$ are defined as
\begin{align}
\label{eq:helical_forms}
	\omega_1 &= dx, \\ 
	\notag
	\omega_2 &= \cos(p x) dy - \sin(p x) dz, \\
	\notag
	\omega_3 &= \sin(p x) dy + \cos(p x) dz,
\end{align}
and form an algebra:
\begin{equation}
\label{eq:domega}
 d \omega_1 = 0, \qquad d \omega_2 = - p \, \omega_1 \wedge \omega_3, \qquad d \omega_3 = p \, \omega_1 \wedge \omega_2.
\end{equation}
They are appropriate for describing a helix with pitch $p$ and direction along the \hbox{$x$-coordinate}. 
It is important to note that because the source $\varLambda(x)$ is fixed on the boundary and is not dynamical, it can absorb the momentum of the bulk system. It forces the geometry to adapt and becomes a channel through which momentum flows towards what is essentially an infinite external ``bath''. The effect of the helical source on the bulk metric can be obtained with the following Ansatz
\begin{gather}
\label{eq:metric}
ds^2 = -U(r) dt^2 + \frac{dr^2}{U(r)} + e^{2 v_1 (r)}\big(\omega_1 \big)^2 + e^{2 v_2(r)} \big(\omega_2 \big)^2 + e^{2 v_3(r)}\big(\omega_3 \big)^2 \\
\notag
B = w(r) \omega_2, \qquad A = a(r) dt.
\end{gather}
The Killing vector fields of this space-time contain the three fields:
\begin{equation}
\xi_1 = \partial_z, \qquad \xi_2 = \partial_y, \qquad \xi_3 = \partial_x - y \partial_z + z\partial_y
\label{eq:killingVectors}
\end{equation}
It is clear from these Killing vectors that in the Bianchi-VII space, the $x$-direction is not characterised by translational invariance. It should be noted at this point that this property of the space is genuine and not an artefact of the coordinate choice and therefore cannot be reversed by a coordinate transformation. In other words there exists no coordinate transformation $x^\mu\to \bar{x}^{\mu}$, such that $\xi_i = \partial_{\bar{x}^{i}}$ \footnote{Except if $v_1=v_2=v_3$.}. If that were possible then the only effect would be a rotation of the direction (beside the already translationally invariant ones) along which translations would be conserved and the momentum vector would simply be re-defined.

The specific form of the functions $U,v_i,\omega$ and $a$ follow from solving the Einstein-Maxwell equations. Insisting on an asymptotic approach to AdS$_5$ at the boundary the Ansatz functions should have the following asymptotics:
\begin{align}
\label{eq:asymtotic_metric}
r\rar\infty: \qquad U(r) \sim r^2, \qquad v_i(r)\sim \ln(r).
\end{align}
In the absence of a TSB potential when $\lambda=0$ (but at non-zero $\mu$) the solution should reproduce the conventional non-extremal Reissner-Nordstr\"om black hole in AdS (RN), and the functions reduce to
\begin{gather}
\label{eq:AdS-RN}
RN: \qquad U(r) = r^2 \left[1 - \left(1 + \frac{\mu^2}{3 r_h^2} \right) \Big(\frac{r_h}{r} \Big)^4  + \frac{\mu^2}{3 r_h^2} \Big(\frac{r_h}{r} \Big)^6 \right], \qquad v_i(r) = \ln(r), \\
\notag
a(r) = \mu \left[1 - \Big(\frac{r_h}{r}\Big)^2 \right], \qquad w(r) = 0.
\end{gather}

At finite $\lambda>0$ the generic solution remains a non-extremal black hole and the temperature is defined by the surface gravity at the horizon:
\begin{equation}
T = \frac{U'(r_h)}{4 \pi},
\end{equation}
where the radius $r_h$ is the largest root of $U(r_h)=0$. Accordingly, the near-horizon asymptotics of the Ansatz read
\begin{equation}
\label{eq:IR_background}
r\rar r_h: \qquad U(r)=(r-r_h)U_h, \quad a(r)=(r-r_h)E_h, \quad w(r)=w_h, \quad v_i(r) = v_i^h.
\end{equation}

The great advantage of the Bianchi-VII model is that despite the fact the metric is explicitly dependent on $x$-coordinate and breaks translational symmetry, the non-linear equations of motion can still be recast in the form of ordinary differential equations (ODEs) in a single radial coordinate \cite{Donos:2012js}, which is why it has been extensively used in the study of various phenomena caused by explicit translational symmetry breaking. In \cite{Donos:2012js} it was shown that the momentum dissipation leads to the appearance of finite resistivity and, moreover, the conductivity can become irrelevant in the IR as one continuously tunes the source $\lambda$. In \cite{Erdmenger:2015qqa} the superconducting phase transition in the presence of a helical lattice was studied, and in \cite{Andrade:2015iyf} the background was analysed from the point of view of commensurability phenomena. Even though the equations of motion are ODEs, finding the background solution for a given $\mu, \lambda, p, T$ represents a considerable numerical computation task. For further details we refer the reader to the aforementioned papers. In this paper we make use of the numerical procedures described in \cite{Andrade:2015iyf} and \cite{Erdmenger:2015qqa}, in parallel, allowing us to cross-check our numerical results.
\begin{figure}[ht]
\centering
\begin{subfigure}[]{0.35\linewidth}
\includegraphics[width=1\linewidth]{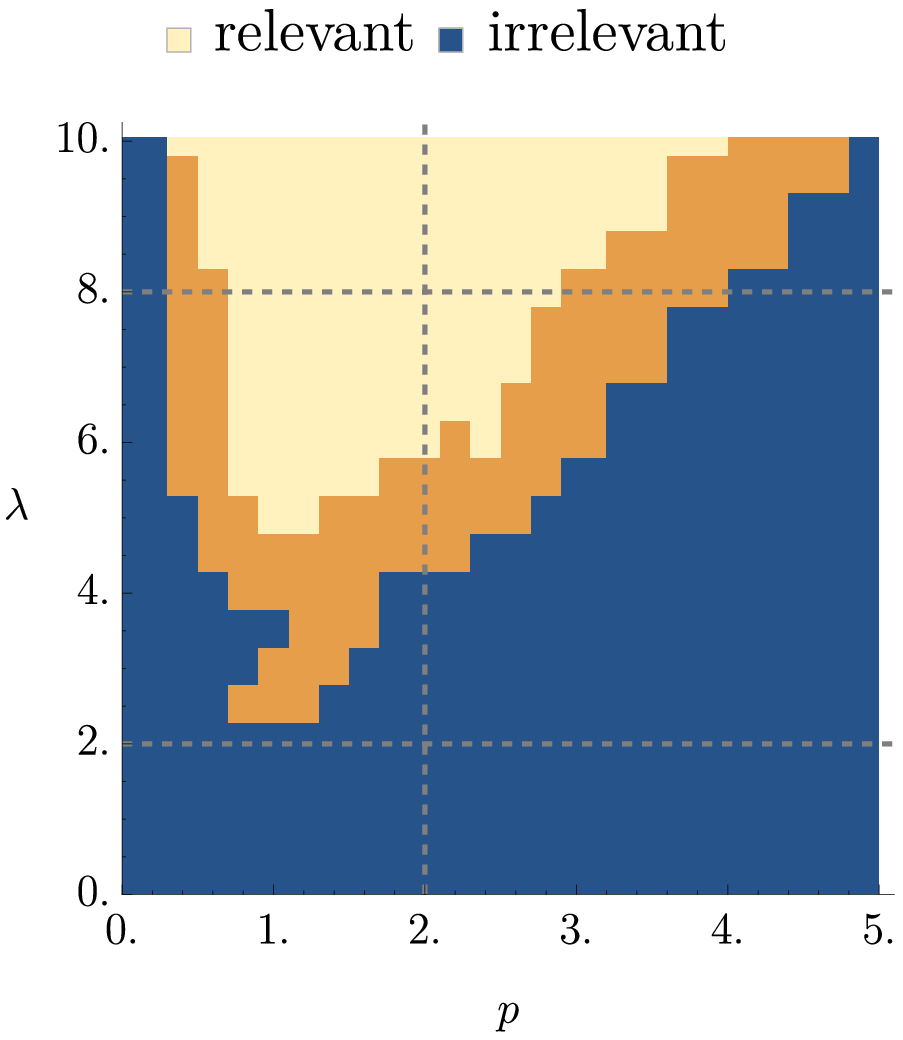}
\end{subfigure}
\qquad
\begin{subfigure}[]{0.5\linewidth}
\includegraphics[width=1\linewidth]{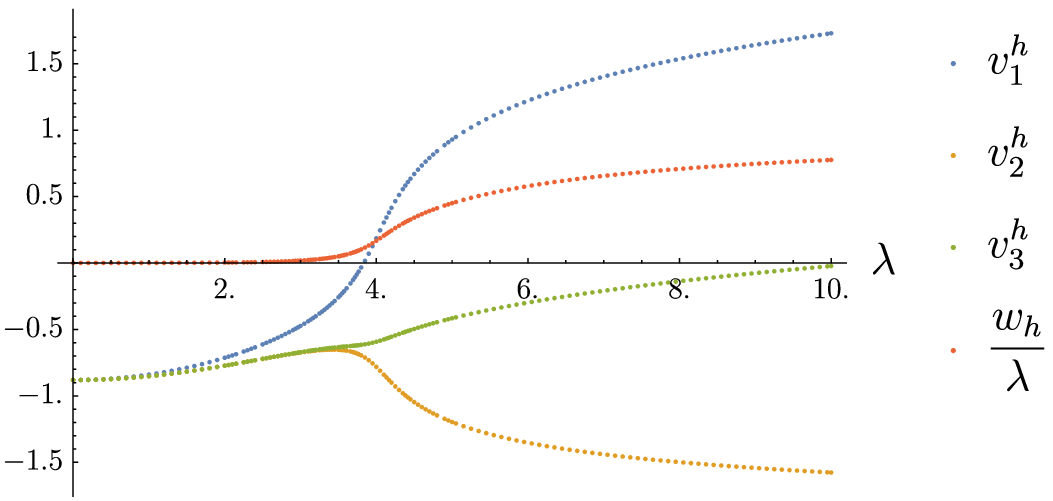}
\end{subfigure}
\caption{\label{fig:phase_diagram} Irrelevant/relevant TSB phase diagram of helical background (left panel). The horizon values of the profile functions \eqref{eq:IR_background} along the cut $p=2$ (right panel) show clearly the crossover around $\lambda \approx 4$ at low temperature $T=1/(80 \pi)$.}
\end{figure}  

The relevant physics of this model is as follows. In \cite{Donos:2012js} it was shown that depending on the parameters of helical deformation $(\lambda, p)$ the model exhibits a quantum phase transition at zero temperature. In the bulk this difference of phases corresponds to a change in thermodynamically preferred near horizon geometry. For a weak helix $\lambda \ll \mu$ one is in a conventional metallic phase with weak TSB that is irrelevant in the IR. That is $w(r_h)$ vanishes and the near-horizon geometry is not modified by the helix. Hence it acquires the same form as in the pure critical RN case, asymptoting to $AdS_2\times \mathbb{R}^3$ at $T=0$
\begin{align}
\label{eq:irrelevant_IR}
\mbox{Irrelevant TSB:}  \qquad & U=12 \epsilon^2, \qquad v_i = v_0, \qquad a = 2 \sqrt{6} \epsilon, \qquad w = 0,  \\
\notag
&T=0, \qquad \mbox{as}\,\, \epsilon \equiv (r-r_h) \rar 0.
\end{align}
For a strong helix the deformation is relevant and the system flows to a new fixed point. Now $w(r_h) \neq 0$ and the near-horizon geometry is modified acquiring the form
\begin{align}
\label{eq:relevant_IR}
\mbox{Relevant TSB:}  & \qquad U=u_0 \epsilon^2,  \qquad a = a_0 \epsilon^{5/3}, \qquad w = w_0 + w_1 \epsilon^{4/3}, \\
\notag
& \qquad e^{v_1} = e^{v_1^0} \epsilon^{-1/3}, \qquad e^{v_2} = e^{v_2^0} \epsilon^{2/3}, \qquad e^{v_3} = e^{v_3^0} \epsilon^{1/3}.
\end{align}
Given the reasoning in the Introduction we will name these two fixed points as {\em irrelevant} and {\em relevant} TSB deformations\footnote{They were named ``metallic'' and ``insulating'' in \cite{Donos:2012js}}.

Each $T=0$ fixed point has a finite-$T$ generalisation satisfying \eqref{eq:IR_background} and we should stress here that we always study such finite albeit small-$T$ configurations. Nevertheless at sufficiently small but finite temperature one can clearly discern the difference by looking at the horizon values of the profiles (see the right panel of Fig.\,\ref{fig:phase_diagram}). As it was shown in \cite{Donos:2012js, Rangamani:2015hka,Erdmenger:2015qqa,Andrade:2015iyf} and as we will confirm below, the qualitative features of the IR physics are in essence governed by the $T\rar 0$ fixed point.  
The phase diagram shown on the left panel of Fig.\,\ref{fig:phase_diagram}, was obtained in \cite{Andrade:2015iyf} by studying the scaling of DC conductivity at small temperature. As we discussed in the Introduction, in the relevant TSB phase the conductivity vanishes at $T\rar0$, while it approximates a Drude peak in the irrelevant TSB case. 
\section{Dirac equation}
\label{sec:Dirac_equation}
We proceed by introducing the minimally coupled fermion field in the helical background. The Dirac equation has the standard form \cite{Faulkner:2009wj,Cubrovic:2011xm}
\begin{equation}	
\label{eq:Dirac}
\left[ \mathbf{e}^\M_A \Gamma^A (\p_\M + \frac{1}{4} \omega_\M^{BC} \Gamma_{BC} - i q A_\M) - m \right] \Psi = 0,
\end{equation}
where $A,B \in \{\ut, \ux, \uy, \uz, \ur \}$ are the tangent space indices (denoted with an over-dot), $\Gamma^A$ are 5D gamma-matrices, $\Gamma^{BC} = \frac{1}{2}[\Gamma^B,\Gamma^C]$, $\mathbf{e}_A^\M$ is the vielbein and $\omega_\M^{BC}$ is the associated spin connection. $q$ is the charge of the fermion and its corresponding operator in the dual field theory; the mass $m$ encodes the scaling dimension $\Delta_\Psi=(m-d/2)$. The vielbein can be chosen in many different ways, so it is natural to bring it to the form possessing the same symmetries as the background metric. The Ansatz \eqref{eq:metric} suggests using the helical 1-forms \eqref{eq:helical_forms} for its definition. We will be using the co-frame
\begin{align}
\label{eq:coframe}
\mathbf{e}^A = \Big(U(r)^{1/2} dt, \, e^{v_1(r)} \omega_1, \, e^{v_2(r)} \omega_2, \, e^{v_3(r)} \omega_3, \, U(r)^{-1/2} dr \Big). 
\end{align}
This co-frame has the neat property that due to \eqref{eq:domega} the spin connection $\omega_A^{BC} = \mathbf{e}_A^\M \omega_\M^{BC}$ does not depend on $x$. In what follows it will also be important that the frame components $\mathbf{e}_{\ut}$ and $\mathbf{e}_{\ux}$ do not depend on $x$ either. 

At this point we can clearly see how the non-conservation of momentum in this background as manifested through its Killing vectors (\ref{eq:killingVectors}) is encoded in the fermionic dynamics. Recall that the current corresponding to translations is $T^{\M\N}\propto \bar{\Psi}\Gamma^\M \nabla^\N \Psi$ and this current is conserved if the system is translationally invariant $\nabla_\M T^{\M\N}=0$. Since fermions are defined on the tangent space, the energy-momentum tensor should be properly re-written for any non-trivial space, in the form $T^{\M\N}\propto \bar{\Psi}\mathbf{e}^\M_A \Gamma^A \nabla^\N \Psi$. The divergence of the $x$-momentum current is therefore proportional to
\begin{equation}
\nabla _\M T^{\M x} = \nabla_\M \bar{\Psi}\mathbf{e}_A^\M \Gamma^A \nabla^x \Psi \propto \bar{\Psi}\mathbf{e}_A^\M \Gamma^A R_{\M x B C} \Gamma^{BC}\Psi ,
\label{eq:fermionStressEnergyTensor}
\end{equation}
where $R_{\M\N AB}$ is the Riemann tensor. For a homogeneous, $x$-translationally invariant space-time like AdS or black-hole AdS it vanishes identically, because the metric components only depend on the radial coordinate. In the Bianchi-VII background space-time however the metric also has $x$-dependence, \eqref{eq:fermionStressEnergyTensor} does not vanish and $x$-momentum is not conserved. This property underpins all of the results that we will present, as it describes the exact way through which momentum is relaxed, along the $x$-direction.

Before continuing we choose a representation of the gamma-matrices. In 5D a spinor has 4 components and the set of gamma-matrices can be obtained from a 4D set $\Gamma^A=\gamma^A$ for $A \neq \ur$ by adding $\Gamma^{\ur}= \pm \gamma^5$, where $\gamma^5= i^{-1}\gamma^{\ux}\gamma^{\uy}\gamma^{\uz}\gamma^{\ut}$ is the usual chiral gamma matrix in 4D \cite{polchinski1998string}. Apparently, there are 2 distinct ways of completing the 5D basis, differing by the choice of the sign of $\Gamma^{\ur}$. One can understand this degeneracy by recalling that due to the absence of chirality in 5D, the holographic fermion in the representation of a given Clifford algebra describes a fermionic operator of the corresponding chirality on the boundary. In other words, a certain choice of the Clifford algebra in the 5D bulk gives us one half of the degrees of freedom of the fermion on the boundary.\footnote{See i.e. \cite{Hong:2006ta} for the relevant discussion in a different set-up of AdS/QCD.} Therefore, in order to describe a Dirac fermion in 4D we need to use both ways of defining gamma-matrices in the bulk, one for each chirality.\footnote{In holographic studies it is usually assumed that one can obtain the results for opposite chirality by switching the sign of the mass term. This is indeed equivalent to the sign change of $\Gamma^{\ur}$ up to some trivial redefinitions of the spinor components, as can be seen from \eqref{eq:Dirac_equations}.}

Since the structure of the background that we are considering has a definite helicity, we would like to have control over projections of the fermionic spin on the direction of the helix which is proportional to $\Gamma^{\uy\uz}$ \cite{Herzog:2012kx}, and we choose the set of gamma-matrices which renders it diagonal. At the same time we wish to have a diagonal $\Gamma^{\ur}$ matrix in order to facilitate the further treatment of the near-boundary asymptotics of the bulk fermionic fields. Both goals can be achieved by choosing (consistent with \cite{Faulkner:2009wj})
\begin{align}
\label{eq:gamma-matrices}
\Gamma^{\ut} & = \begin{pmatrix} i \sigma_1 & 0 \\ 0 & i \sigma_1 \end{pmatrix}, &
\Gamma^{\ux} & = \begin{pmatrix} \sigma_2 & 0 \\ 0 &  - \sigma_2 \end{pmatrix}, \\   
\notag
\Gamma^{\uy} & = \begin{pmatrix} 0 & - \sigma_2 \\  - \sigma_2 & 0 \end{pmatrix}, & 
\Gamma^{\uz} & = \begin{pmatrix} 0 &  - i \sigma_2 \\  i \sigma_2 & 0 \end{pmatrix}, &
\Gamma^{\ur} & = \pm \begin{pmatrix} -\sigma_3 & 0 \\ 0 & -\sigma_3 \end{pmatrix}
\end{align}
where the choice of ``+'' or ``-'' in $\Gamma^{\ur}$ leads to the left or right chirality of the boundary fermion respectively, as discussed above. As desired, the $x$-angular momentum is given by the diagonal operator
\begin{equation}
\label{eq:sigma_yz}
\Gamma^{\uy\uz} = \begin{pmatrix} -i \mathbb{1} & 0 \\ 0 & i \mathbb{1} \end{pmatrix},
\end{equation}
and we can split the 4-component spinor $\Psi$ into two $2$-component Weyl spinors with given spin projections denoted as:
\begin{equation}
\Psi = (\psi_\uparrow, \psi_\downarrow)^T
\end{equation}
Thanks to the choice of fermion representation, the Dirac equation \eqref{eq:Dirac} for different spin components decouples completely. After adopting the frame \eqref{eq:coframe} we can rescale the fermionic field to eliminate the spin connection from the equations of motion and, using the aforementioned fact that the coefficients do not depend on $x$, perform a Fourier transform. 

Before turning to the momentum representation it is useful to make the following observation. The frame \eqref{eq:coframe} has a helical structure and ``rotates'' synchronously with the background. It helps in simplifying the equations, but is also confusing because one has to work in a rotating basis. In the limiting case when $\lambda = 0$ the helix disappears and the background reduces to AdS-RN \eqref{eq:AdS-RN}, but this frame does not reduce to the trivial ``static'' frame in the boundary Minkowski space. In order to study the boundary fermions in this ``static'' frame one needs to unwind them by applying the $x$-dependent Lorentz rotation. Due to the fact that the rotation generator $\Gamma^{\uy\uz}$ is diagonal \eqref{eq:sigma_yz}, the corresponding unwinding of a spinor $\Psi \rar \mathrm{exp}(- i p x \Gamma^{\uy\uz}/2) \Psi$ reduces to a simple shift in momentum space by $p/2$ and $-p/2$ for opposite spins, respectively.

Taking all these ingredients into account, we arrive at the transformation
\begin{equation}
\Psi(x_\mu,r) = \exp\left[-i \omega t + i \left(k_x \mp \tfrac{p}{2} \right) x + i k_y y + i k_z z \right]  \big(U e^{2(v_1 + v_2 + v_3)} \big)^{-1/4} \Phi(r).
\end{equation}
where $``\mp" \rar -1$ for spin up components and $``\mp" \rar +1$ for spin down. The Dirac equation for right chirality takes the following form:
\begin{multline}
\label{eq:Dirac_equations}
\left[ \p_r - \frac{m}{\sqrt{U}} \sigma_3 - \frac{e^{-v_1}}{\sqrt{U}} \left( \frac{p}{2} [\mathrm{cosh}(v_2-v_3)-1] \mp k_x \right) \sigma_1 + \frac{\omega + q A_t}{U} \ i \sigma_2 \right] \phi_{\uparrow,\downarrow}^R \\
+ \frac{e^{-v_2}}{2 \sqrt{U}} \Big[\left(1 + e^{v_2-v_3}\right) (k_y \pm i k_z) + \left(1-e^{v_2-v_3}\right) e^{\mp 2 i p x} (k_y \mp i k_z) \Big] \sigma_1  \phi_{\downarrow,\uparrow}^R = 0, 
\end{multline}
where the choice of signs is associated with the up and down spin states, respectively. The equations for left chirality have opposite signs in front of the $\p_r$ derivatives. 

Let us now analyse the structure of the Dirac equation in the Bianchi-VII TSB background. The only place were the non-homogeneity of the background appears in the equations is the $x$-dependent part of the coupling between opposite spins. This coupling generically leads to the Umklapp scattering with momentum transfer $2 p$ between the modes in different Brillouin zones, giving rise to the corresponding band gaps. Interestingly though, the coupling is also proportional to the value of the projection of the fermion momentum, which is transverse to the helical axis -- $k_y,k_z$ in this case. As also explained in the Introduction this is a distinctive feature of the helical ``homogeneous lattice'' background. Indeed, the helix preserves the diagonal combination of the translation symmetry along $O_x$ and rotation symmetry in the transverse plane. Until the probe, propagating in this geometry, singles out a certain direction in the $(O_y,O_z)$-plane, the diagonal symmetry subgroup allows for the homogeneous, $x$-independent, dynamics. Any finite transverse momentum of the probe breaks the rotation symmetry alongside with the translations along $O_x$. The Umklapp coupling should of course vanish in the limit $\lambda=0$. This is easily verified as in that case one has the RN background \eqref{eq:AdS-RN} with $v_2(r)=v_3(r)$.    

Another interesting feature of equations \eqref{eq:Dirac_equations} becomes apparent when one sets $k_y=k_z=0$. Even though the $x$-dependent part vanishes, the helical structure of the background is still manifested through the chiral asymmetry. In the $p=0$ case the equation for spin ``up'' right fermion coincides with the spin ``down'' left fermion. In order to demonstrate this, one needs to transpose the components of the $\phi_\downarrow^L$ 2-spinor and use the symmetry of the equations under the simultaneous flip of the spin and change of sign of $k_x$. When $p \neq 0$ this matching is not possible any more due to the change of sign in front of $p$. This behaviour is indeed expected because the background has a definite helicity and thereby breaks P-symmetry. Henceforth the left and right fermions are not degenerate.

As it has already been argued in the Introduction, we can qualitatively illustrate the above features of bulk holographic fermions by looking at the dual boundary theory. Given the bottom-up nature of this holographic construction we cannot directly derive an explicit action for the boundary field theory, but we can employ the symmetries of the bulk equations in order to deduce the possible lowest order terms in the effective Lagrangian. From the point of view of the boundary one can think about a low-energy effective theory in which the chiral components of the Dirac spinors don't mix. We should rather consider a pair of Weyl fermions, which we denote by $\theta$. The helical source $\varLambda(x)=\lambda \omega_2$, which is present on the boundary as an external field, gives rise to an effective potential $V[\lambda]$. This potential consists of at least two parts: one of them accounts for the transverse-momentum-dependent Umklapp coupling, discussed above, and the other is independent of $x$ and lifts the degeneracy between chiralities due to the broken P-symmetry. 

The Umklapp term must meet the following requirements: It is proportional to the transverse momentum of the fermion, hence it must include the derivative operator, and it is periodic with momentum $2 p$, including two instances of the helical source $\varLambda(x)$, and it preserves chirality. This is enough to fix the form of this term as $V_{\mathrm{Umklapp}} \sim \gamma^\mu \varLambda_\mu \varLambda^\nu \p_\nu$.

As to the homogeneous part of the potential, the obvious choice would be $V = V(\Lambda^2)$, but this will not be enough as it doesn't posses the information about the helicity of the background and will only shift the energy levels uniformly. The other $x$-independent value is an ``angular momentum'' pseudo-vector $\Omega^\mu=\epsilon^{\mu \nu \rho \tau} \p_\nu \varLambda_\rho \varLambda_\tau$. This must be multiplied by the pseudo-vector fermionic current which preserves chirality $\bar{\theta} \gamma_\mu \gamma^5 \theta$. Thus we can illustrate the effective dynamics of the low-energy degrees of freedom by the following action
\begin{align}
\label{eq:boundary_action}
S_{4D} = \int d^4x \ \bar{\theta} \left[\gamma^\mu \p_\mu + \left(\mu + \varkappa_1 |\varLambda|^2 \right) \gamma^0 + \varkappa_2 \gamma^5 \gamma^{\mu} \Omega_\mu  + \varkappa_3 \gamma^\mu \varLambda_\mu \varLambda^\nu \p_\nu \right] \theta,
\end{align}
with some phenomenological coupling constants $\varkappa_i$. The second term accounts for the presence of the chemical potential. For a given form of $\varLambda = \lambda \omega_2$ the pseudo-vector $\Omega$ has only the $t$-component $\Omega = p \lambda^2 dt$, hence the effect of the $\varkappa_2$ term reduces to the chirality-dependent shift of the effective chemical potential, which is similar to the magnetic field induced splitting of levels in semi-metals. The equations of motion obtained from this action have a qualitative form, discussed in the Introduction \eqref{eq:boundary_equation}. There is one crucial difference between the effective action \eqref{eq:boundary_action} and the exact holographic system. The action \eqref{eq:boundary_action} is minimally coupled to the helical lattice and does not dissipate, in the usual way. This is due to lattice momentum conservation. In the true holographic system the lattice is non-minimally coupled through the vielbein. Because of this, momentum can be transferred to the helical background, which acts as a perfect momentum sink.

\section{Analytic treatment}
\label{sec:analytic}
\subsection{Schr\"odinger potential}
\label{sec:potential}
As has been pointed out in \cite{Cubrovic:2011xm, Faulkner:2009wj,Hartnoll:2012rj,Iqbal:2011ae} much of the fermion physics in a holographic set-up can be qualitatively understood by examining the effective Schr\"odinger potential which arises when one rewrites the equations of motion in a second order form. Generically the form of the potential is described by the three features: (a) near the AdS boundary (the UV in the dual field theory) the potential rises, corresponding to the fact that all (massive) particles in AdS can never reach the boundary, (b) near the horizon it becomes strongly negative and describes the fermions falling through the horizon, corresponding to the finite lifetime of the boundary field theory excitations, (c) when the bulk fermion mass (boundary scaling dimension) is sufficiently small a potential barrier develops well into the intermediate region, where quasi-bound states can form. These bound states correspond to the holographic quasi-particles. These can tunnel through the barrier to the horizon, and this is reflected in the self-energy of these holographic quasi-particles. For the pristine RN black hole, this barrier becomes infinitely high precisely at zero energy and at Fermi momentum $k_F$, resulting in the emergence of infinitely long-lived quasi-particles, defining a precise Fermi-surface. (d) For large Fermion masses, this barrier does not exist and excitations can only tunnel into the horizon. This tunnelling rate encodes the spectrum of these excitations and is proportional to the near-horizon AdS$_2$ propagator $\mathcal{G}(k, \omega) \sim \omega^{2\nu_k}$. The holographic quasi-particles are formed with propagators $G (\omega, k)^{-1} = \omega - v_{\scalebox{0.6}{F}} k  - \Sigma (\omega, k)$ where $\Sigma \sim \omega^{2\nu_{k_{\scalebox{0.4}{F}}}}$, with the caveat that $\Sigma'' (\omega=0, k=k_F) = 0$. 

As an example, we take the equations for the spin up right chirality component of \eqref{eq:Dirac_equations}. One can rewrite the system of two first order ODEs as one second order ODE for a single component of the Weyl spinor $\phi_\uparrow^R = (\phi_1,\phi_2)^T$. Performing an additional coordinate transformation it is brought to the Schr\"odinger form $[\p_s^2  - V(s)]\phi_1=0$ with the potential
\begin{equation}
\label{eq:schroedinger_equation}
V(s)=- \frac{1}{c_0^2 (K-\Omega)^2}\left(- M^2  - K^2 + \Omega^2 - \p_r M + \frac{\p_r(K-\Omega)}{K-\Omega} \right), 
\end{equation}
where the new coordinate is $s = c_0 \int_{r_0}^{r} dr (K-\Omega)$. The $s$-dependent profiles in the case of ($R\uparrow$)~fermion are
\begin{gather}
M = \frac{m}{\sqrt{U}}, \qquad K = \frac{e^{-v_1}}{\sqrt{U}}\left(\frac{p}{2}[\mathrm{cosh}(v_2-v_3)-1] - k\right), \qquad \Omega = \frac{\omega + q A_t}{U}. 
\end{gather}
For any finite frequency $\omega$ the range of $s$ is unbounded (we discuss the special case of $\omega=0$ in the Appendix \ref{app:zero_frequency}). Due to the near-horizon behaviour of $\Omega \sim \omega/(r-r_h)$, the integral for $s$ diverges logarithmically. Therefore it is convenient to set $r_0$ to the UV radius $r_\infty$, so that the AdS boundary is now located at $s=0$ and the horizon is at $s \rar -\infty$. In this region the potential \eqref{eq:schroedinger_equation} flattens out and approaches the finite value $V\sim -c_0^{-2}$.

The form of the fermionic response is now critically dependent on the existence of bound states with zero energy in this potential well. The frequency and the momentum of the bulk fermion are parameters of the potential. The presence of a bound state, a solution with vanishing boundary conditions at a given $(\omega,k)$, also known as a quasi-normal mode (see Sec.\,\ref{sec:quasinormal_modes}), is associated with a pole in the spectral function of the holographically dual field theory (see Sec.\,\ref{sec:spectral_function}). A zero energy bound state is in direct correspondence to the existence of well defined quasi-particles in the spectrum of the dual boundary theory.
\begin{figure}
\centering
\includegraphics[width=1. \linewidth]{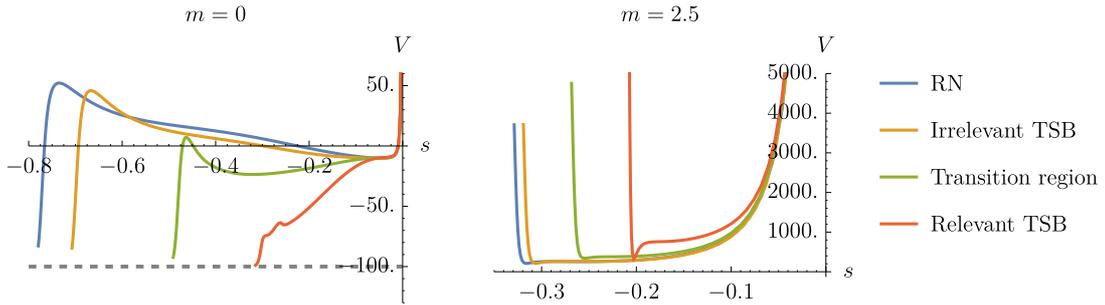}
\caption{\label{fig:potentials_finite_omega} Form of the Schr\"odinger potentials \eqref{eq:schroedinger_equation} in various regimes at zero energy ($\omega=0.001$).  We choose $\lambda=2$ for irrelevant TSB  phase, $\lambda=6$ for relevant TSB and $\lambda=4$ in transition region; RN corresponds to $\lambda=0$. The boundary is at $s=0$ whereas the horizon at $s\to-\infty$. We set $k=0.8\approx k_F$ for $m=0$ plot, and $k=0$ for $m=2.5$. $c_0=10$ and $p=2$.}
\end{figure}
The set of potential profiles for fermion mass $m=0$ is shown in the left plot of Fig.\,\ref{fig:potentials_finite_omega}. We assume a finite, but very small value of $\omega$ and the momentum $k$ is chosen so that the $\lambda=0$ profile crosses the zero-energy level, exhibiting a zero-energy bound state. In other words, this momentum coincides with the Fermi momentum $k_F$ in the RN set-up. For $\lambda=0$ the bound state is separated from the horizon at $s\rar -\infty$ by a large potential barrier rendering the quasi-particle state very long-lived. Upon turning on the translational symmetry breaking source $\lambda>0$ we observe that this barrier gets suppressed in the near-horizon IR region and is eventually washed out. At moderate $\lambda$ the quasi-particle at $\omega=0$ will acquire a finite life-time while eventually at large $\lambda$, in the relevant TSB phase where the IR geometry is qualitatively different, the bound state will disappear, becoming a runaway wave-function. Hence, in the exact spectral function we expect that the quasi-particle states fade away, to eventually completely disappear from the fermionic response functions.

The other case of interest is when the fermion mass gets so large that the potential does not cross the zero energy level at any $k$, so that the quasi-particle bound states disappear even in the RN set-up \cite{Faulkner:2011tm,Cubrovic:2011xm}. For example the corresponding profiles for $m=2.5$ are shown on the right plot of Fig.\,\ref{fig:potentials_finite_omega}. No substantial change is observed in the potential profile when one turns on the TSB helical background; bound states are not formed. Accordingly, we do not expect drastic changes in the fermionic response in this regime.
\subsection{IR Green function}
\label{sec:IRGreenFunction}
This Schr\"odinger potential analysis is particularly useful to isolate the presence of bound states corresponding to sharp excitations in the spectrum. The full properties of this low-frequency fermionic response function can in fact be understood as the combination of two contributions. The presence or absence of this bound state and the contribution from a local quantum critical IR fixed point\cite{Faulkner:2009wj,Iqbal:2011ae,Faulkner:2011tm,Zaanen:2015oix}. This one is fundamentally unrelated to quasi-particles. In the case of large mass, when the quasi-particle contribution is absent, the fermion response is dominated by this non-Fermi liquid part. It is possible to reconstruct the boundary response by studying the near-horizon geometry of the model and performing the ``matching trick'' at small $\omega$ \cite{Faulkner:2009wj,Faulkner:2011tm,Iqbal:2011ae}. The fermionic Green function is therefore related to the ``IR Green function'', computed in the near horizon geometry. In the case of zero temperature the IR Green function has a simple power-law form: $\mathcal{G}|_{T=0} = \omega^{2\nu_k}$, with a momentum dependent power $\nu_k$. On the complex $\omega$-plane this gives rise to a branch cut, originating from $\omega=0$. At finite temperatures the branch-cut resolves into a series of poles which we therefore name the ``thermal poles'' of the Green function. From the asymptotic form of the Dirac equations, one can show (see Appendix \ref{app:IR_green} for details) that at finite temperature the contribution of the IR fermion Green function to the density of states at small $\omega$ has the same form as in \cite{Faulkner:2011tm}\footnote{Our calculation coincides with (5.23) of \cite{Faulkner:2011tm} up to a choice of signs in the first term, which proves to be irrelevant.}:
\begin{equation}
\label{eq:G_IR}
\mathcal{G}_R = (4 \pi T)^{2 \nu} \frac{(m-i \tilde{m}) \tilde{R}_2 - i q \tilde{e}_d - \nu}{(m-i \tilde{m}) \tilde{R}_2 - i q \tilde{e}_d + \nu} \cdot \frac{\Gamma(-2 \nu) \Gamma(\tfrac{1}{2} + \nu - \frac{i \omega}{2 \pi T} + i q \tilde{e}_d) \Gamma(1 + \nu - i q \tilde{e}_d)}{\Gamma(2 \nu) \Gamma(\tfrac{1}{2} - \nu - \frac{i \omega}{2 \pi T} + i q \tilde{e}_d) \Gamma(1 - \nu - i q \tilde{e}_d)},
\end{equation}
where the constants get corrections due to the helical background \eqref{eq:IR_background}:
\begin{gather}
\label{eq:IR_constants}
\tilde{R}_2 = \left(\frac{7 E_h^2 - 24}{12} + p^2 \, \frac{\mathrm{sh}\big[v_2^h - v_3^h\big]^2}{e^{2 v_1^h}} + \frac{5}{12} p^2 \frac{w_h^2}{e^{2(v_2^h+v_3^h)}} \right)^{-2}, \qquad \tilde{e}_d = E_h \tilde{R}_2^2 \\
\notag
\tilde{m} = \frac{1}{\sqrt{U_h} e^{v_1^h}} \left[k - \frac{p}{2} \left(\mathrm{cosh}(v_2^h - v_3^h) - 1\right) \right], \qquad  \nu =  \sqrt{\tilde{R}_2^2 (m^2 + \tilde{m}^2) - q^2 \tilde{e}_d^2}.
\end{gather}
One discerns the set of ``thermal'' poles, which is given by the zeroes of the $\Gamma$-function in the denominator.

In the metallic phase the near-horizon limit of the helical background at zero temperature is $AdS^2\times \mathbb{R}^3$ \eqref{eq:irrelevant_IR}. In this case the corrections in \eqref{eq:IR_constants} vanish due to the fact that $v_2^h=v_3^h$ and $w_h=0$. Therefore the spectral density at low $\omega$ behaves in the same way as the one in the RN case, even though the helical source in the UV is not vanishing. On the other hand, the near-horizon geometry of the relevant TSB phase is affected by the helical lattice operator (see Fig.\,\ref{fig:phase_diagram}). At zero temperature it is substantially different from $AdS^2\times \mathbb{R}^3$ \eqref{eq:relevant_IR}. In the case of zero temperature the IR equations of motion have an irregular singular point at $r=r_h$ and it is not clear to us whether an analytic expression for $\mathcal{G}_R$ exists. Nevertheless at finite temperature the role of the helical deformation is to provide sizeable corrections to \eqref{eq:IR_constants}, while the formula itself should still hold. In the following Sections we will compare the prediction \eqref{eq:G_IR} with the numerical results and check the reliability of the near-horizon treatment in both cases.
\section{Spectral function}
\label{sec:spectral_function}
With this intuition at hand, we are now ready to proceed with solving the Dirac equations numerically to obtain the retarded two-point function of the boundary fermionic operator dual to the bulk fermion field $\Psi$. We will mostly be interested in the spectral function which is defined as the trace over spin polarizations of the imaginary part of the retarded Green's function:
\begin{equation}
\label{eq:spectral}
\rho(\omega, k) = -\frac{1}{\pi} \mathrm{Tr}\, \mathrm{Im} G^R (\omega, k).
\end{equation}
The retarded Green's function can be obtained in the standard way \cite{Iqbal:2009fd,Faulkner:2009wj,Cubrovic:2009ye,Cubrovic:2011xm,Liu:2009dm, Iqbal:2011ae,Zaanen:2015oix} by solving the bulk equations of motion \eqref{eq:Dirac_equations} with in-falling boundary conditions at the black hole horizon 
\begin{align}
\label{eq:BC_IR_Green}
\phi_\alpha \sim (r-r_h)^{- i \frac{\omega}{U_h}}+\dots,
\end{align}
and a fixed infinitesimal source at the AdS boundary. Once the profile for the bulk fermionic field is known, the source and the response of the dual boundary operator are encoded in the coefficients of leading and sub-leading terms of its near-boundary expansion. Substituting the asymptotic background values \eqref{eq:asymtotic_metric} into the equations of motion \eqref{eq:Dirac_equations} we obtain the expansion for, e.g., right chirality spin up component $\phi_\uparrow^R = (\phi_1,\phi_2)^T$. The leading terms are:
\begin{align}
\label{eq:BC_UV_Green}
(\uparrow, R): \qquad \begin{pmatrix}
\phi_1 \\ \phi_2
\end{pmatrix} = \begin{pmatrix}
A(\omega,k) (r^m+\dots) + B(\omega,k) (r^{-m -1}+\dots) \\
C(\omega,k) (r^{m-1}+\dots) + D(\omega,k)( r^{-m}+\dots)
\end{pmatrix}
\end{align}
and similar expansions can be derived for the other polarizations. The dots refer to sub-leading terms in $r^{-1}$. In practice we use up to 6 orders of the expansion of UV boundary conditions: the details are listed in Appendix\,\ref{sec:AppendixQNMs}. The corresponding holographic response function of this spinor component is associated with the ratio\footnote{$B(\omega,k)$ and $C(\omega,k)$ coefficients do not bear any additional information since due to the first order nature of the Dirac equation $\phi_1$ and $\phi_2$ components do not represent independent degrees of freedom.}
\begin{equation}
\mathcal{S}_{\uparrow R, \uparrow R} = \frac{D(\omega,k)}{A(\omega,k)}.
\end{equation}
The total response function on the boundary is, of course, a $4\times4$ matrix in the spinor representation and every component of the spinor with certain chirality and spin contributes to one entry on its diagonal. The off-diagonal components vanish due to the choice of the gamma matrices \eqref{eq:gamma-matrices}, making it possible to completely decouple the equations of motion. The retarded Green's function can then be obtained as a suitable rotation of the response matrix:\footnote{See \cite{Iqbal:2009fd} for details.}
\begin{equation}
G^R = - i \mathcal{S} \gamma_0.
\end{equation}

The complete spectral function \eqref{eq:spectral}, which is a sum of four terms corresponding to different fermionic polarisations, is obtained by numerically solving the equations of motion \eqref{eq:Dirac_equations} with boundary conditions \eqref{eq:BC_UV_Green}, \eqref{eq:BC_IR_Green} for every value in the $(\omega, k)$ domain in the given helical background characterized by two parameters: $(\lambda, p)$. The numerical calculations are performed for two values of the fermion mass discussed in Sec.\ref{sec:analytic}: $m=0$ and $m=2.5$. All dimensionful quantities are normalized with respect to the chemical potential $\mu$. The calculations are performed for finite, but low temperature
\begin{equation}
\label{eq:temperature}
T = \frac{1}{80 \pi} \mu.
\end{equation}
With this in hand we can follow how the structure of collective excitations in the dual boundary theory changes when the system undergoes the irrelevant to relevant TSB transition, by examining the spectral function. We scan the parameter space $(\lambda, p)$ along the trajectories shown on Fig.\ref{fig:phase_diagram}. The scan with fixed $p=2$ and $\lambda \in [0,10]$ shows an evolution of the spectral function upon the irrelevant-relevant phase transition. The two scans with fixed $\lambda$ allow us to explore the dependence of the results on the helical pitch $p \in [0,5]$ in the irrelevant TSB ($\lambda=1$) and relevant TSB ($\lambda=6$) phases. We will now discuss the results.

\begin{figure}[ht]
\centering
\includegraphics[width=1 \linewidth]{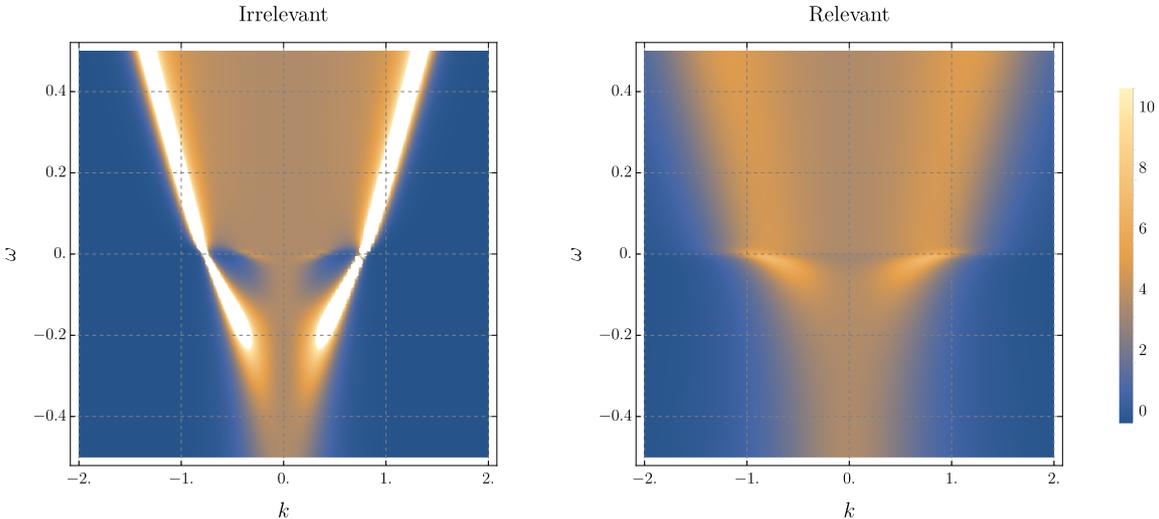}
\caption{\label{fig:spectral_met-ins}The spectral function for $m=0$ as measured in the irrelevant TSB and relevant TSB phase. The former corresponds to $\lambda=1, p=2$, the latter to $\lambda=6, p=2$.}
\end{figure}

\begin{figure}[ht]
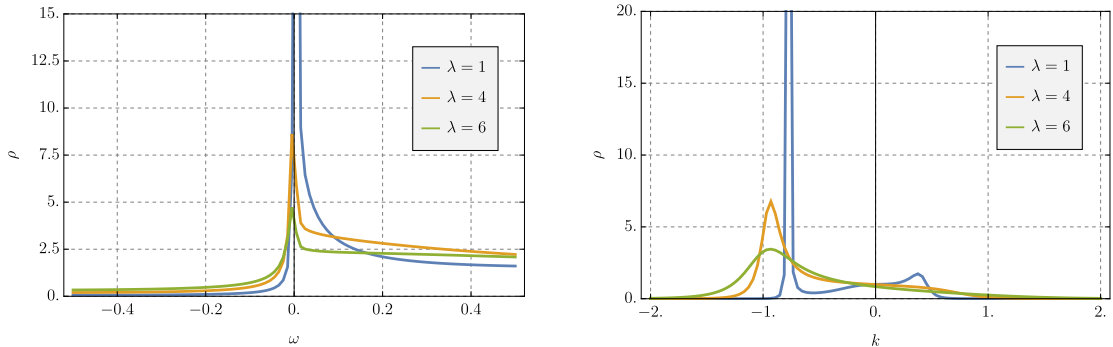

\centering
\begin{subfigure}{0.45\linewidth}
\includegraphics[width = 1\linewidth]{{{figs/helicalDensityCutVsw}}}
\label{fig:DensityCutsw}
\end{subfigure}
\qquad
\begin{subfigure}{0.45\linewidth}
\includegraphics[width = 1\linewidth]{{{figs/helicalDensityCutVsk}}}
\label{fig:DensityCutsk}
\end{subfigure}
\caption{Density of states $\rho$, for $k=k_F$ as a function of $\omega$ (left) and for $\omega=0$  as a function of momentum $k$ (right). The lines correspond to increasing helical strength, corresponding to the irrelevant ($\lambda=1$), intermediate ($\lambda=4$) and relevant ($\lambda=6$) phases.}
\label{fig:DensityCuts}
\end{figure}

For the low fermion mass $m=0$, everywhere in the irrelevant TSB phase one can see the pronounced Dirac cone and a well defined Fermi surface near $\omega=0$, related to the presence of quasi-particles (see left plot on Fig.\,\ref{fig:spectral_met-ins} and corresponding cuts on Fig.\,\ref{fig:DensityCuts}). This is indeed expected from our qualitative analysis of the Schr\"odinger potential of Sec.\,\ref{sec:potential}. We see, that even though the momentum dissipation is sizeable in the irrelevant phase, as long as the IR geometry remains $AdS^2\times\mathbb{R}^3$ the quasi-particles remain well defined degrees of freedom in the dual theory. These quasi-particles are expected to behave in accordance to the effective boundary field theory \eqref{eq:boundary_action} sketched in Sec.\,\ref{sec:Dirac_equation}. In particular, one should be able to detect the asymmetry between the two chiralities, which leads to the splitting of the Fermi momenta. We evaluated the components of the spectral function separately for different chiralities and studied this asymmetry. For the parameters we have been using, the splitting is of order $\Delta k_F/k_F \sim 10^{-4}$ and is barely visible.\footnote{For instance when $\lambda=2, p=2$ we get $k_F^{R} = 0.7994, k_F^{L}=0.7995$. For larger $\lambda$, when $\lambda=6, p=4.2$ the splitting is more pronounced:  $k_F^{R} = 0.9657, k_F^{L}=0.9665$} One would expect this effect to be more pronounced in the region of the parameter space with larger $\lambda$ and $p$.

The behaviour of the $m=0$ fermion spectral density in the relevant TSB state is very difficult (see right plot of Fig.\,\ref{fig:spectral_met-ins}). One sees immediately that the quasi-particles have disappeared. The spectral function is heavily damped and it's weight is suppressed -- there is barely any remnant of the sharp Dirac cone. As we stated in the Introduction this has nothing in common with a conventional insulator, characterized by the reconstruction of the fermion dispersion relation and opening of the gap in the band structure, no sign of a gap is observed. Instead it appears that the quasi-particles themselves become ill defined. This is exactly according to expectations based on the analysis of the Schr\"odinger potential. 

There is a higher spectral density near the former Fermi momentum points around $\omega = 0$ (see also Fig.\,\ref{fig:DensityCuts}). This is quite unnatural from the point of view of conventional condensed matter theory. The spectral density on the Fermi surface must be either zero, or display a sharp peak related to the quasi-particles, with the width fixed by the temperature. At zero temperature the quasi-particles in a Fermi liquid are infinitely sharp on the Fermi surface regardless of the strength of the interaction. In the case under consideration one clearly sees that the quasi-particles acquire sizeable width at $\omega=0$, which is much larger then the temperature scale and signals the deconstruction of the Fermi surface itself due to the strong momentum relaxation in the helical lattice. We will address this issue in more detail by studying the behaviour of the quasi-normal modes, associated with poles in the Green's function, in the next Section.
\begin{figure}[ht]
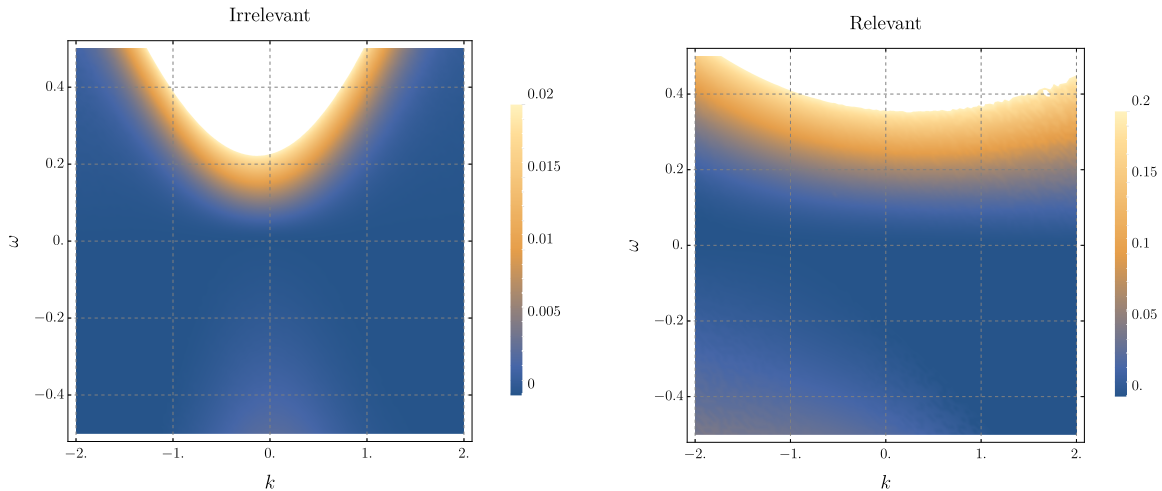

\centering
\begin{subfigure}[t]{0.48\linewidth}
\includegraphics[width=1. \linewidth]{{{figs/DensityPlots25_Metallic}}}
\end{subfigure}
\qquad
\begin{subfigure}[t]{0.46\linewidth}
\includegraphics[width=1. \linewidth]{{{figs/DensityPlots25_Insulating}}}
\end{subfigure}
\caption{\label{fig:density_large_Mass} Spectral function for $m=2.5$ in irrelevant and relevant phase. The contribution from only left chirality, spin up component is shown. The irrelevant phase corresponds to $\lambda=1, p=2$, while the relevant to $\lambda=6, p=2$. No sign of quasi-particles is observed.}
\end{figure}

\begin{figure}[ht]
\centering
\includegraphics[width=0.49 \linewidth]{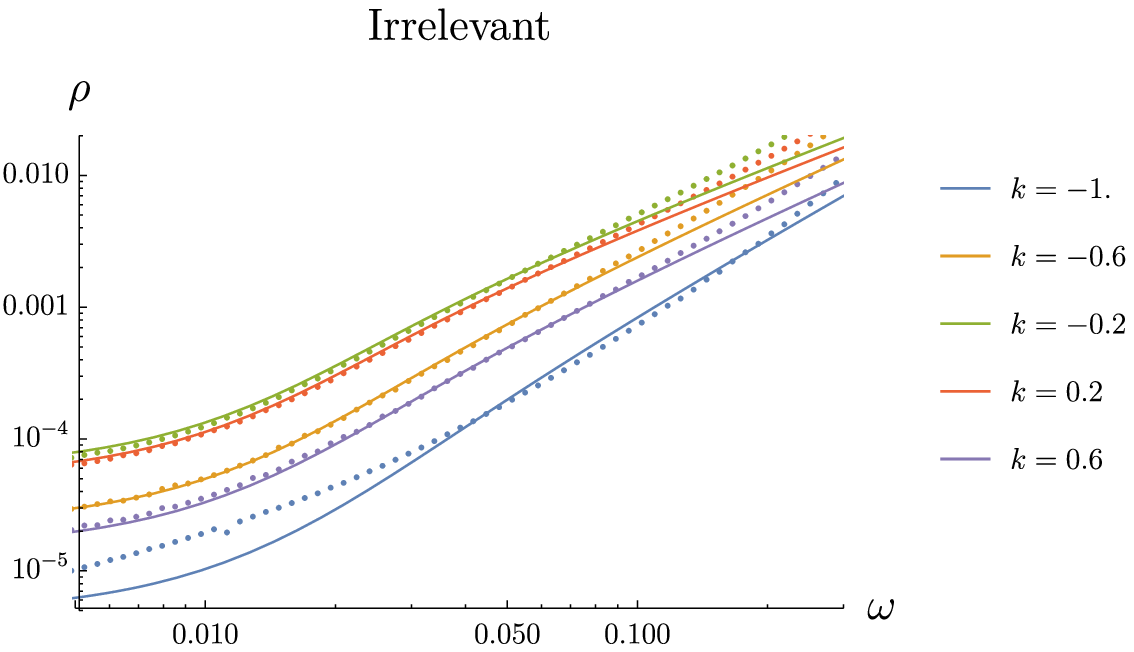}
\includegraphics[width=0.49 \linewidth]{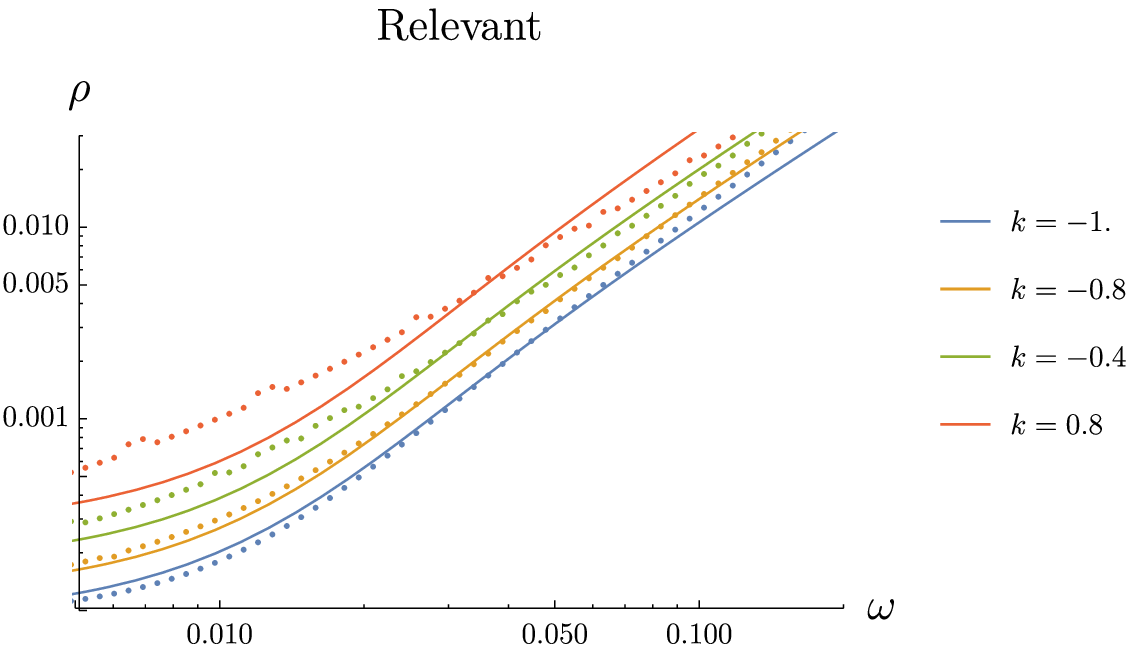}
\caption{\label{fig:density_cuts} Constant-$k$ cuts of the spectral function for $m=2.5$: irrelevant and relevant phase. The former corresponds to $\lambda=1, p=2$, the latter to $\lambda=6, p=2$. Solid lines show the analytic prediction of \eqref{eq:G_IR}, whereas the points correspond to the numerical results.}
\end{figure}

We now turn to the case of the large fermion mass $m=2.5$. In accordance to the Schr\"odinger potential study, there are no longer quasi-particle peaks in the spectral functions which are quite featureless both in the irrelevant and relevant states (Fig.\,\ref{fig:density_large_Mass}). It is interesting to examine the constant-$k$ cuts of the spectral functions in order to check the scaling behaviour at low temperatures anticipated from the IR Green function matching, discussed above. These profiles are shown in Fig.\,\ref{fig:density_cuts}. In both cases the data exhibit good agreement in the intermediate scales $10^{-2} < \omega < 10^{-1}$. At large $\omega$ one does not expect the IR matching procedure to give reliable estimates, while in the small $\omega$ region our numerical accuracy is no loner sufficient. One can see, that the excellent matching between the data points and analytic predictions \eqref{eq:G_IR} around $k=0$ in the irrelevant TSB case and $k=1$ in the relevant TSB case is quickly destroyed when we consider wider range of momenta. Importantly, this is happening not only in the relevant TSB regime, where one can expect deviations from the scaling \eqref{eq:G_IR} due to the reconstruction of the low temperature geometry, but also in the irrelevant one, where there are no physical reasons for mismatch. This signals that our numerical results, which are limited by the accuracy of the calculated gravitational background, become less reliable when one considers larger momenta. With the current precision we cannot resolve the breakdown of the analytic IR scaling formula, discussed earlier, in relevant TSB case.  
\section{Quasi-normal modes}
\label{sec:quasinormal_modes}
In the previous section several interesting features of the fermionic dynamics were identified, the most striking being the fading away of the Fermi peaks as the system transitions from the irrelevant TSB to the relevant TSB phase. In order to better understand and quantify these phenomena we will in this section turn to the description of holographic response functions in terms of quasi-normal modes (QNMs). Quasi-normal modes correspond to poles in the Green's function of the operators under consideration. They constitute a particularly convenient tool in holography since they can be readily computed and capture important information of the system under investigation. In this case we are interested in the QNMs of the fermionic Green's function and we will compute them numerically, following the procedure detailed in Appendix~\ref{sec:AppendixQNMs}, which is similar to the computation of the spectral functions in Sec.\,\ref{sec:spectral_function}.

\begin{figure}[!htbp]
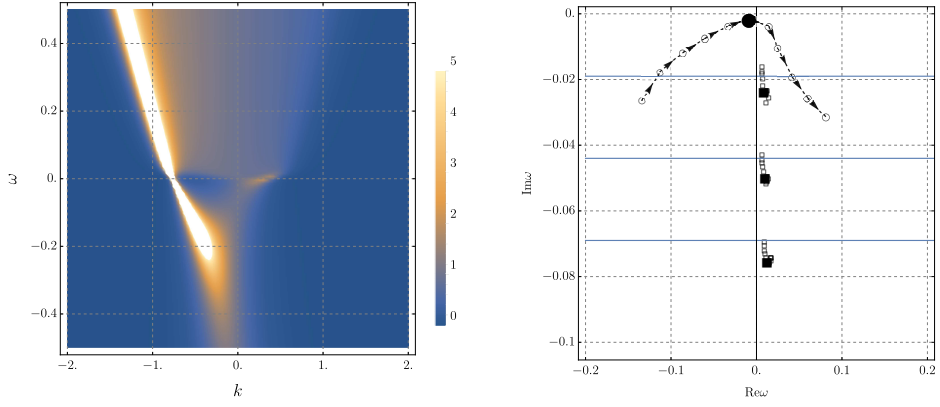
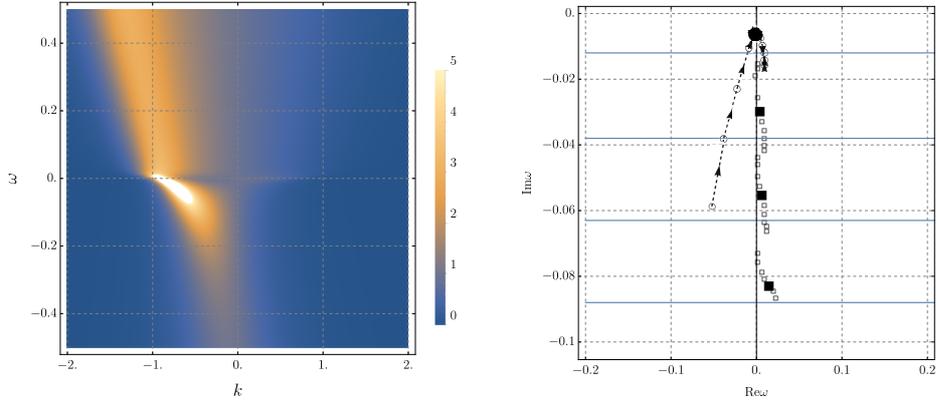
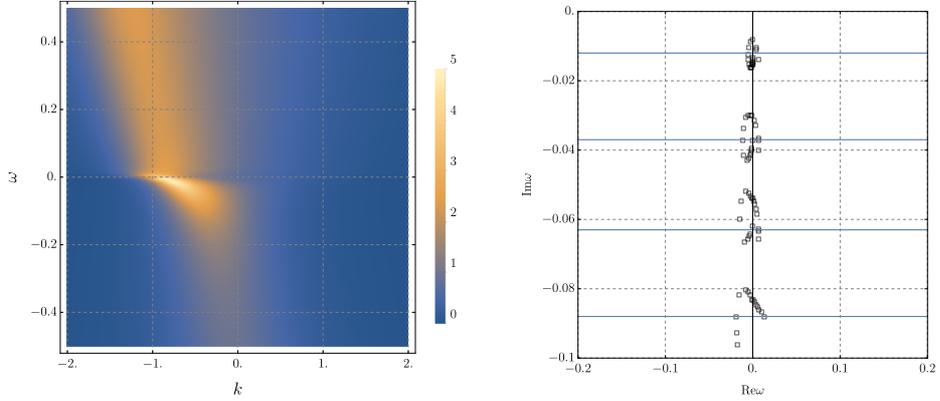

\centering
\begin{subfigure}[t]{0.75\linewidth}
\includegraphics[width = .52\linewidth]{{{figs/helicalDensity-mF=0-lambda=1-p=2}}}
\qquad
\includegraphics[width = .48\linewidth]{{{figs/helicalQNMs-mF=0-lambda=1-p=2}}}
\caption{Irrelevant phase $\lambda=1$. Trajectory of the quasi-particle pole is seen and has the apex at $k=k_F=-0.75$.}
\label{fig:DensityQNMsMetallicA}
\end{subfigure}

\begin{subfigure}[t]{0.75\linewidth}
\includegraphics[width = .52\linewidth]{{{figs/helicalDensity-mF=0-lambda=4-p=2}}}
\qquad
\includegraphics[width = .48\linewidth]{{{figs/helicalQNMs-mF=0-lambda=4-p=2}}}
\caption{Irrelevant-relevant transition region $\lambda=4$. quasi-particle pole acquires significant imaginary part. The apex momentum is $k=-0.9$.}
\label{fig:DensityQNMsTransitionA}
\end{subfigure}

\begin{subfigure}[t]{0.75\linewidth}
\includegraphics[width = .52\linewidth]{{{figs/helicalDensity-mF=0-lambda=6-p=2}}}
\qquad
\includegraphics[width = .48\linewidth]{{{figs/helicalQNMs-mF=0-lambda=6-p=2}}}
\caption{Relevant phase $\lambda=6$. quasi-particle pole can not be identified.}
\label{fig:DensityQNMsInsulatingA}
\end{subfigure}
\caption{The density of states (left panels) and the QNMs (right panels) in the case of $m=0$. Parameters are: $p=2$, $T=\frac{1}{80\pi}$. The positions of QNMs are shown for momentum $k$ ranging from $k=-0.5$ to $k=-1.5$ in the direction of the eye-guides. The blue lines show the estimations for the thermal QNMs coming from \eqref{eq:G_IR}.}
\label{fig:DensityQNMs}
\end{figure}

We first want to establish the understanding of how the QNMs capture the known dynamics of our system in the irrelevant case. In Fig.\,\ref{fig:DensityQNMsMetallicA} a plot of the spectral function, as derived in the previous section, along with the corresponding results for QNMs are shown side by side. For clarity, we focus only on one of the spinor components unlike Fig.\,\ref{fig:spectral_met-ins}, where all four components are combined.

The right panel of Fig.\,\ref{fig:DensityQNMsMetallicA} shows the set of fermionic QNMs for progressively smaller values of momenta in an overlaid fashion. In other words, moving along the horizontal direction in the spectral function on the left panel of Fig.\,\ref{fig:DensityQNMsMetallicA} corresponds to tracing the motion of the QNMs. For each momentum value $k$ we observe two types of QNMs. One category includes the poles (three of which are depicted by squares in Fig.\,\ref{fig:DensityQNMsMetallicA}) that lie on the complex plane close to the imaginary axis, with almost exclusively imaginary parts.  These ``thermal'' poles are characteristic of low-temperature QNMs and can be readily read off from the expression for the IR Green function \eqref{eq:G_IR} (the actual analytic predictions following from \eqref{eq:G_IR} are shown on the plots by blue lines.) 
The other type of QNM is the one that appears to have a significant real part, while for some values of $k$ it is actually the one closest to the real axis. We will call this quasi-particle pole, since it corresponds to the peak in the spectral function near $\omega=0$ which is the one governing the low-energy excitations near the Fermi surface. Equivalently, this is the pole associated with the Weyl cone. As we tune $k$ away from $k_F$, which in Fig.\,\ref{fig:DensityQNMsMetallicA} would correspond to moving along the cone away from the Fermi surface, the pole recedes into the complex plane, acquiring a larger imaginary part and the corresponding quasi-particle therefore begets a finite width at finite frequency. Conversely when we tune $k$ close to $k_F$ we see the quasi-particle pole approaching the origin, which is an indication of long-lived quasi-particle dynamics near the Fermi surface. The apex quasi-particle pole, i.e. the one with minimum imaginary part due to $k\approx k_F$ is emphasised on the right panel of Fig.\,\ref{fig:DensityQNMsMetallicA} by a larger, full circle. Similarly, the set of the thermal poles corresponding to the apex values of $k$ are highlighted by larger, filled squares. 

To summarise the irrelevant case, we observe the standard thermal series of poles, with a moderate dependence on the momentum. On top of these we see the isolated quasi-particle pole, with Weyl-cone dynamics and a sharp $k$-dependence, corresponding to the quasi-particle excitations around the Fermi surface. Departure from momentum values close to $k_F$ results in broadening of said quasi-particles.

Let us now increase the strength of the helix $\lambda$. It is instructive to examine the intermediate $\lambda = 4$ case separately, roughly corresponding to the edge of the irrelevant phase where the transition to the relevant TSB is happening. In Fig.\,\ref{fig:DensityQNMsTransitionA} the QNMs and spectral function are juxtaposed as before. One can see that the quasi-particle pole now acquires a significant imaginary part and its ``trajectory'' when we scan through momentum~$k$ dives more rapidly down the imaginary half-plane. The broadening of the quasi-particle at finite $\omega$ is much stronger than before; the window around the Fermi surface where the quasi-particle can be defined becomes quite narrow. More strikingly, as we have already noticed in the previous Section, the imaginary part of the pole remains sizeable even at the apex, i.e. at the would-be Fermi surface. Hence due to the increased momentum dissipating potential the quasi-particle has a finite lifetime even at the Fermi surface, leading to a finite spectral weight at $\omega=0$.

\begin{figure}[ht]
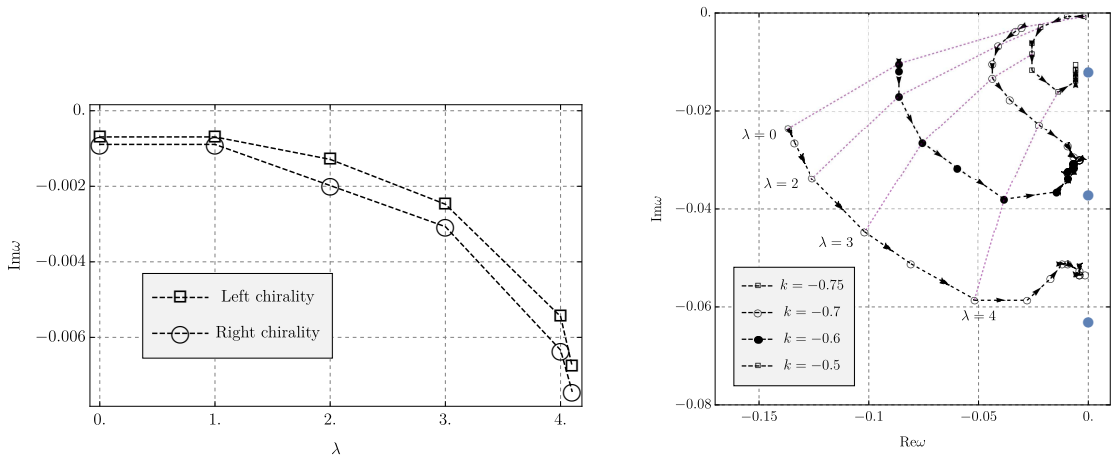

\centering
\begin{subfigure}{0.5\linewidth}
\includegraphics[width=1\linewidth]{{{figs/fermiPoleVsLambdaComb}}}
\caption{The imaginary part of the apex quasi-particle pole for the right and left chirality. The dashed line is a guide for the eye.}
\label{fig:FermiPoleVsLambda}
\end{subfigure}
\qquad
\begin{subfigure}{0.4\linewidth}
\includegraphics[width=1\linewidth]{{{figs/QNMs.k-slices}}}
\caption{Trajectories of the quasi-particle pole for $k=-0.75,-0.7,-0.6,-0.5$ when $\lambda$ is increased from $0$ to $10$. The quasi-particle pole joins the thermal series along the imaginary axis (blue dots estimated from \eqref{eq:G_IR}) at $\lambda \gtrsim 4$.}
\label{fig:QNMskSlices}
\end{subfigure}
\caption{Dynamics of the quasi-particle pole at increasing lattice strength.}
\label{fig:QNMs_analisys}
\end{figure}

We can examine this issue in more detail by following the quasi-particle pole, tuned to the apex $k\simeq k_F(\lambda)$, where it has the minimal imaginary part and dominates the irrelevant phase, for increasing values of $\lambda\in[0,4]$. The results are presented in Fig.\,\ref{fig:FermiPoleVsLambda}. The imaginary part of the quasi-particle pole starts at a non-zero value at $\lambda=0$, as we are at low but non-zero temperature, and then becomes more negative as $\lambda$ increases. The progressively larger size of the imaginary part gives rise to a dissipative behaviour, which results in smearing of the Fermi surface already in the irrelevant phase. Moreover as the helical strength is tuned closer to the transition value, which is roughly $\lambda\simeq 4$, the rate at which the imaginary part of the quasi-particle pole becomes more negative, increases. It must be stressed here that the results so far have been of left chirality, as defined in Eq.~\ref{eq:gamma-matrices}. For completeness in Fig.\,\ref{fig:FermiPoleVsLambda} we present the dependence of the quasi-particle pole on $\lambda$ for the right chirality as well. The observed difference between left and right chiralities are almost insignificant.

Pushing now into the relevant TSB phase, at $\lambda=6$, the picture becomes even more intriguing. The density of states and QNMs in this regime are shown on Fig.\,\ref{fig:DensityQNMsInsulatingA}. Here we see only the thermal poles, being consistent with the theoretical prediction \eqref{eq:G_IR} at a given temperature and $\lambda$. No significant dependence on the momentum $k$ is observed any longer, meaning that there is no distinct $k_F$ where quasi-particle poles govern dynamics. Reflecting the non-existence of a distinguished pole, in Fig.\,\ref{fig:DensityQNMsInsulatingA} all of them are depicted identically. The four apparent bunches are essentially the poles at different values of $k$. As we see, for large values of $\lambda$, into the relevant TSB phase, the notion of a quasi-particle pole becomes ill-defined, which is consistent with the fast decrease of the imaginary part in Fig.\,\ref{fig:FermiPoleVsLambda}. 

Before proceeding further we would like to better understand the transition between the irrelevant and relevant phases as reflected in the QNMs and in particular in the ``fate'' of the quasi-particle pole. For this purpose we sampled a set of fixed momenta and followed the associated poles for increasing values of $\lambda$. The results are show in Fig.\,\ref{fig:QNMskSlices}. It is clear from this plot that for low values of $\lambda$ the by now familiar arches tracing the quasi-particle pole emerge. In other words for small $\lambda$ one can tune the momentum so that the quasi-particle pole comes close to the origin and therefore governs the low-energy excitations. In the transition region, near $\lambda\simeq 4$, the slope with respect to the momentum becomes steeper as $\lambda$ increases and eventually the poles that we used to name quasi-particle poles hit the imaginary axis and become indistinguishable from the other thermal poles predicted by \eqref{eq:G_IR}. At this point distinction between quasi-particle and other poles becomes meaningless. The particular endpoint of the evolution of the quasi-particle pole depends on the momentum, but is always next to a certain member of the thermal series. In particular, at $k=-0.75$ which for our set of parameters is the Fermi momentum at $\lambda=0$ (i.e. in absence of momentum dissipation), it evolves into the lowest lying thermal pole. This behaviour reflects a substantial reconstruction of the fermionic Green's function at the irrelevant-relevant phase transition. While in the irrelevant case the characteristics of the Green's function could be attributed to the ``quasi-particle'' and ``quantum critical'' part, in the relevant TSB phase they are merged into a new, un-factorizable form, described by the new discrete series of poles in the complex plane. We should also note here that the series of thermal poles at $\lambda=6$ is substantially different from the thermal poles at $\lambda\simeq 0$, because the deviations in the parameters \eqref{eq:IR_constants} are sizeable.
\begin{figure}[ht]
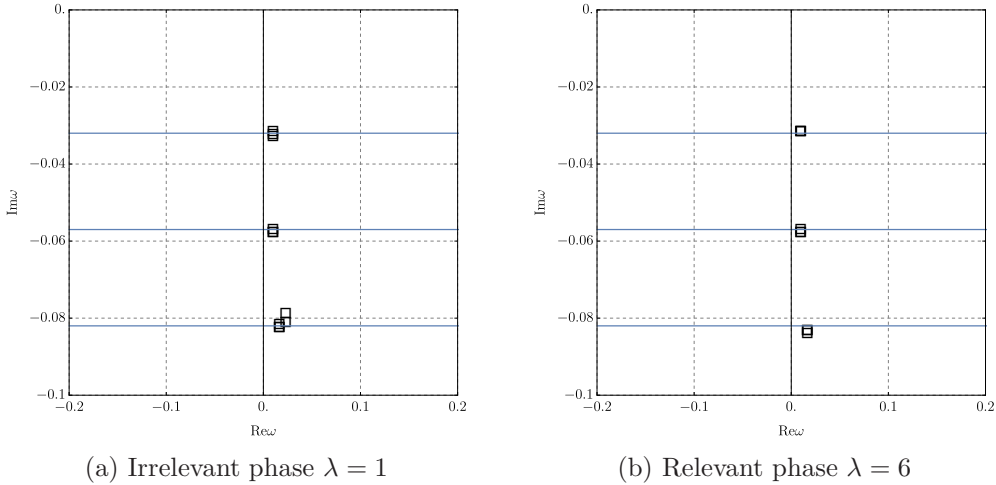

\centering
\begin{subfigure}[t]{0.4\linewidth}
\includegraphics[width = 1.\linewidth]{{{figs/helicalQNMs-mF=2.5-lambda=1-p=2}}}
\caption{Irrelevant phase $\lambda=1$}
\label{fig:MetallicMassiveFermionDensityQNMs}
\end{subfigure}
\qquad
\begin{subfigure}[t]{0.4\linewidth}
\includegraphics[width = 1.\linewidth]{{{figs/helicalQNMs-mF=2.5-lambda=6-p=2}}}
\caption{Relevant phase $\lambda=6$}
\label{fig:InsulatingMassiveFermionDensityQNMs}
\end{subfigure}
\caption{The QNMs in the case of large mass $m=2.5$. Parameters are $p=2$ and $T=\frac{1}{80\pi}$.}
\label{}
\end{figure}

Finally we turn towards the strange metal regime by increasing the mass of the fermions in which case there are no Fermi surfaces to begin with as discussed in the previous Section. The expectations from the spectral function calculation is that there should be no low-lying QNMs, since we observe no spectral weight near the $\omega\simeq 0$ region. As before we compute the corresponding QNMs first in the irrelevant phase Fig.\,\ref{fig:MetallicMassiveFermionDensityQNMs} and subsequently in the relevant TSB phase Fig.\,\ref{fig:InsulatingMassiveFermionDensityQNMs}. Indeed we observe that in both irrelevant and relevant phases there are no low-lying QNMs. Instead we see only poles down the imaginary axis, which coincide with the poles coming from the analytic form of the IR Green function \eqref{eq:G_IR}. There is no qualitative difference between the irrelevant and relevant state, confirming our previous claim that in the case of large mass the dynamics in both phases is governed by the IR geometry equally well.
\section{Conclusion}
\label{sec:conclusion}
In this work we have studied how strong translational symmetry breaking (TSB) affects fermionic correlators with the use of holography. The gravitational model we used is the Bianchi-VII (helical) background. Translational symmetry breaking affects the fermions in two characteristic ways. When the momentum transverse to the helix does not vanish $k_\perp\neq 0$ Umklapp scattering occurs, based on an effective action that we deduced, resembling the phenomenology of Condensed Matter systems. For the special case $k_\perp =0$, however, no Umklapp effect can be observed even though momentum conservation is violated along the $x$-direction, i.e. that of the helical axis. 

In both cases translational symmetry breaking is different from well-known lattice physics in condensed matter. There the lattice breaks translational invariance down to a discrete symmetry ($\mathbb{Z}$), hence Umklapp occurs but the Bloch momentum is effectively conserved within the Brillouin zone. In the Bianchi-VII model translational symmetry while still periodic is broken completely, in the sense that momentum conservation is violated at all scales in addition to Umklapp.

Choosing $k_\perp=0$ throughout this work, we are focusing on this latter kind of momentum relaxation which is different to what is observed in condensed matter systems. This mechanism of translational symmetry breaking effectively consists of an external, infinite bath coupled to our system and which can absorb momentum. The role of this classical sink is to externally source translational symmetry breaking potential. Notice that the geometry of the Bianchi-VII space is the channel through which momentum is transferred. The TSB potential is encoded in a field that in turn forces the geometry to take the form that allows momentum relaxation. 

The Bianchi-VII model has two phases distinguished by the (ir)relevancy of the TSB potential. In the irrelevant TSB regime at small $\lambda$,  we observe the existence of well-defined quasi-particles. We have shown that these develop finite life-times, or in other words the quasi-normal pole that corresponds to them acquires a non-zero imaginary part, for any finite value of the helical strength $\lambda$. This is in striking disagreement with regular Fermi-liquid theory, since the latter is expected to be a stable fixed point, immune to small perturbations, whereas here we see that we can continuously attenuate the supposedly infinite-lived quasi-particles. Given our analysis this should come as no surprise -- the background turns momentum into a progressively worse quantum number and therefore quasi-particles, which should carry definite momentum, cannot survive and become damped. This can also be seen through a Schr\"odinger potential analysis of the holographic Dirac equation that governs fermionic excitations. At small $\lambda$ it supports bound states, whereas it becomes shallower as $\lambda$ increases until fermions fall unimpeded through the horizon, exhibiting purely relaxational behaviour. Yet another way of understanding this comes from the symmetry discussion above -- since momentum conservation is violated at all scales in the Bianchi-VII model, even small perturbations around the Fermi surface are not protected.

For values of the helical strength where the Bianchi-VII model transitions to relevant TSB, the picture changes significantly. In this regime there are no quasi-particles left and the Green's function is dominated by ``thermal'' poles, that match very well the analytic prediction based only on the infra-red geometry. Similarly, for large-dimension operators  in the strange-metal regime of the Bianchi-VII background either for irrelevant or relevant TSB, no quasi-particles can be found and the IR calculation perfectly describes the correlator. This is consistent with momentum not being a meaningful quantum number any more. The strange-metal spectral function is rather insensitive to the momentum -- all the quantities that have none or weak momentum dependence remain unaffected by the violation of its conservation.

The study of the Bianchi-VII background using finite-momentum probes and in particular fermions opens up an interesting perspective. Since it can accommodate both Umklapp and non-Umklapp momentum conservation violation, it can be used as a laboratory to explore the interplay between the two phenomena. An interesting balance can be noted here -- the mechanisms of breaking translational symmetry most familiar to CMT are the most challenging in holography though they are within reach, while the easiest holographic mechanism turns out to be quite novel for CMT. At the same time this study demonstrates, that if one intends to describe holographically lattices that are relevant for condensed matter applications, one needs to consider periodic lattices instead of homogeneous ones. Given the aforementioned understanding on the origin of translational invariance breaking, based on the remaining symmetries, we believe that the last statement is quite generic, beyond the specifics of the helical model.

\acknowledgments
We appreciate the contribution of Steffen Klug and Napat Poovuttikul at the beginning stages of this project. We would like to thank Mikhail Katsnelson, Aristomenis Donos, Matthias Kaminski, Christopher Rosen and Tomas Andrade for valuable discussions.

This work is supported in part by the VICI grant of K.S. from the Netherlands Organization for Scientific Research (NWO), by the Netherlands Organization for Scientific Research/Ministry of Science and Education (NWO/OCW) and by the Foundation for Research into Fundamental Matter (FOM). The work of A.K. is partially supported by RFBR grant 15-02-02092a. N.K. is supported by a grant from the John Templeton foundation. The opinions expressed in this publication are those of the authors and do not necessarily reflect the views of the John Templeton foundation. 

This work was carried out on the Dutch national e-infrastructure with the support of SURF Foundation.
\appendix
\section{Numerical methods}
\label{sec:AppendixQNMs}
In this Appendix we will elaborate on the numerics used for the calculations of this paper. Most of what follows is shared between the calculation for the spectral density and QNMs. For both calculations we need the asymptotic expansion of the fermionic fields near both the boundary and the horizon. These expansions are computed by iteratively solving the Dirac equation in the respective region and they depend on knowledge of the corresponding expansions of the background fields. The latter are similarly computed by recursively solving the background equations of motion near the boundary and the horizon. This is necessary because otherwise the background fields are only known numerically. The background expansions are relatively computationally cheap and we were able to push them up to sixth order for the IR and twelfth for the UV, which were more than enough.

With these background field expansions at hand we then proceeded to expand the fermionic fields. First we need to impose the in-falling boundary condition. At first order the near-horizon asymptotic behaviour of the background fields is:\footnote{Here we set $r_h=1$ by employing the symmetry of the equations of motion associated with the overall shift of the radial coordinate: $r \rar r+ \delta r$.}
\begin{equation}
U=U^h(r-1),\,\, v_i=v^h_i,\,\, A_t=E^h(r-1).
\label{eq:BackgroundIR}
\end{equation}
Substituting \ref{eq:BackgroundIR} into the Dirac equation \eqref{eq:Dirac_equations} for, e.g., spin up right chirality, we obtain:
\begin{eqnarray}
(U^h-1)\partial_r \phi_1-\omega \phi_2=0,\\
(U^h-1)\partial_r \phi_2+\omega \phi_1=0.
\end{eqnarray}
Introducing the variable $\epsilon=r-1$ and rewriting the equations in a second order form:
\begin{equation}
\partial^2_\epsilon\phi_1+\frac{1}{\epsilon}\partial_\epsilon \phi_1+\frac{\omega^2}{U_h^2\epsilon^2}\phi_1=0,
\end{equation}
resulting in the near boundary asymptotics of the fermionic component:
\begin{equation}
\phi_1\sim (r-1)^{\pm i\frac{\omega}{U_h}},
\end{equation}
where the ``minus'' sign corresponds to the in-falling boundary condition. The other component at the horizon is then equal to
\begin{equation}
\phi_2=-i\phi_1.
\end{equation}
Similar boundary conditions can be imposed for other spin projections and chiralities. We then recursively solve the Dirac equation up to sixth order in the UV (which turned out to be an overkill) but more importantly up to third order in the IR. The IR expansion demanded some more care -- what is meant by third order is
\begin{equation}
\psi \simeq \sum_{i} c_i (r-1)^i ,\  \  i=0,\frac{1}{2},1,\frac{3}{2},2,\frac{5}{2},3.
\end{equation}
The half-integer powers are required by the nature of the singular points of the equations of motion \eqref{eq:Dirac_equations}. Expansion only in integer powers would result in an algebraic system of equations for the coefficients $c_i$, that has no solution. Another, equivalent, way would be to re-define the background fields that appear under a square root in the Dirac equation. It is important to note that increasing the IR order turned out to be necessary, in the case of the QNMs, in order to be able to distinguish the ones down the imaginary axis. Initial attempts to limit the order to the absolute minimum, obscured that part of the complex plane, whereas the eventual choice allowed us to reach all the way to the numerical limit imposed by the precision with which the background was computed.

Having computed the UV and IR expansions for the fermionic fields we can now determine boundary conditions for the bulk fermions. In the IR we only have one degree of freedom left, since we have imposed in-falling boundary conditions, therefore there is only one linearly independent solution, which we determine by numerically integrating the corresponding initial-value problem. In the UV there are two independent solutions, as indicated by the two integration constants that are left un-determined by the iterative solution. We construct these solutions by imposing two sets of linearly independent boundary conditions, encoded in the choice of the integration constants. In order to compute the spectral function we perform both UV and IR integrations up to an intermediate point, where we demand they match. In fact we solve
\begin{equation}
\alpha \cdot \psi_{UV, \uparrow\downarrow}^{1} + \delta \cdot \psi_{UV, \uparrow\downarrow}^{2} = \psi_{IR, \uparrow\downarrow}
\end{equation}
for $\alpha$ and $\delta$, where the indices $1,2$ refer to the two linearly independent solutions. From the standard fermionic ``entries'' of the holographic ``dictionary'', we know that given the UV asymptotics
\begin{align}
\psi_{\uparrow} \simeq A \cdot r^{m} + \ldots + B\cdot r^{-m-1} + \ldots\\
\psi_{\downarrow} \simeq C \cdot r^{m-1} + \ldots + D \cdot r^{-m} + \ldots
\end{align}
the Green's function is $\propto \frac{D}{A}$, which in our scheme is simply $\frac{\delta}{\alpha}$. From there the spectral function is trivially calculated, for a range of frequencies $\omega$ and momenta $k$.

In the case of QNMs, by definition, we need to impose one extra boundary condition, namely the vanishing of the source on the boundary. This requirement eliminates one of the two linearly independent solutions stemming from the UV and we only need to integrate once from the boundary and once from the horizon, meeting at an intermediate point where matching is tested by requiring
\begin{equation}
\det \left. \left( \begin{array}{cc}\psi_{IR,\uparrow} & \psi_{UV,\uparrow}\\ \psi_{IR,\downarrow} & \psi_{UV,\downarrow}\end{array}\right)\right|_{r=r_M} = 0,
\end{equation}
where $r_M$ is the matching point. The complex frequency plane is then scanned for these zeros, which correspond, by construction, to the QNMs.

Of course, given the introduction of an artificial scale, namely the matching point $r_M$, all the results must be and have been tested for independence against said scale, by repeating the calculations for a range of matching points without observing any significant change.
\section{IR Green function} 
\label{app:IR_green}
At finite temperature we can make use of the asymptotic value of the background profile \eqref{eq:IR_background} and rewrite it in a form: 
\begin{gather}
U(r)\Big|_{r\rar r_h} = \tilde{R}_2^{-2}(r-r_h)(r + r_h - 2 \tilde{r}_{*}), \\
R_2 = \left(\frac{7 E_h^2 - 24}{12} + p^2 \, \frac{\mathrm{sh}\big[v_2^h - v_3^h\big]^2}{e^{2 v_1^h}} + \frac{5}{12} p^2 \frac{w_h^2}{e^{2(v_2^h+v_3^h)}} \right)^{-2} , \quad \tilde{r}_{*} =r_h -\frac{\tilde{R}_2^2}{2} U_h.
\end{gather}
this allows us to introduce the variables
\begin{equation}
\zeta = \frac{\mathfrak{s} \tilde{R}_2^2}{r-\tilde{r}_{*}}, \qquad \zeta_0 = \frac{\mathfrak{s} \tilde{R}_2^2}{r_h-\tilde{r}_{*}}, \qquad \tau = \mathfrak{s} t, \qquad \tilde{\omega} = \mathfrak{s}^{-1} \omega
\end{equation}
and, by taking $\mathfrak{s} \rar \infty$, express the background \eqref{eq:metric} in the near-horizon region as
\begin{gather}
ds^2 = \frac{\tilde{R}^2_2}{\zeta^2} \left( - f(\zeta) d\tau^2 + \frac{d \zeta^2}{f(\zeta)}\right) + e^{2 v_i^h}\big(\omega_i \big)^2, \qquad f(\zeta) = 1 - \frac{\zeta^2}{\zeta^2_0}, \\
A_\tau = \frac{\tilde{e}_d}{\zeta} \left(1 - \frac{\zeta}{\zeta_0}\right), \qquad \tilde{e}_d = E_h \tilde{R}_2^2.
\end{gather}
Finally, we bring the equations of motion \eqref{eq:Dirac_equations} to a form identical to the one used in \cite{Faulkner:2011tm}:
\begin{align}
\left( \p_\zeta - i \hat{\sigma}^3 \frac{\tilde{\omega} + q A_\tau}{f}\right) \tilde{\Phi} = \frac{\tilde{R}_2}{\zeta \sqrt{f}} \left(m \hat{\sigma}^2 + \tilde{m} \hat{\sigma}^1 \right) \tilde{\Phi},
\end{align}
where $\hat{\sigma}^i = \mathbb{1}_2 \otimes \sigma^i$ and the new spinor field is $\tilde{\Phi} = \frac{1}{\sqrt{2}}(1+i \hat{\sigma}^1) \Phi$. The value of $\tilde{m}$ is modified by the helical profile as well:
\begin{gather}
\tilde{m} = \frac{1}{\sqrt{U_h} e^{v_1^h}} \left[k - \frac{p}{2} \left(\mathrm{cosh}(v_2^h - v_3^h) - 1\right) \right].
\end{gather}
After these identifications are made we can use the result of \cite{Faulkner:2011tm} in order to write down the expression for the retarded IR Green function at small $\omega$ and finite temperature \eqref{eq:G_IR}.
\section{Schr\"odinger potential at zero frequency}
\label{app:zero_frequency}
Let us analyse the special case of $\omega=0$ in the Shr\"odinger potential \eqref{eq:schroedinger_equation}. Due to the near-horizon behaviour \eqref{eq:IR_background} in this case the $s$-coordinate spans a finite interval which can be rescaled to unity by appropriate choice of $c_0$ and assuming $r_0=r_h$:
\begin{equation}
\omega=0: \qquad \frac{1}{c_0} =  \int_{r_h}^{\infty} dr (K-\Omega)
\end{equation}
Moreover, one can study the value of the potential exactly on the horizon $s = 0$. Using the expansions \eqref{eq:IR_background} we find
\begin{equation}
\label{eq:IR_potential}
V(s)\Big|_{s\rar 0} = \frac{1}{c_0^2} \frac{\tilde{k}^3 + m^2 \tilde{k}  - \frac{q}{2} E_h m }{\tilde{k}^3}, \qquad \tilde{k} = e^{-v_1^h} \left(k +  \frac{p}{2} \left[1 -\mathrm{ch}(v_2^h - v_3^h)\right] \right).
\end{equation}
One can see that due to the exponential factor $e^{-v_1^h}$ the helical background reduces the effective value of the momentum on the horizon, which leads eventually to the negative value of the potential (see Fig.\,\ref{fig:phase_diagram}). This effect is consistent with the momentum dissipating nature of the considered background.
\bibliographystyle{JHEP}
\bibliography{helical_fermions}
\end{document}